\documentclass[twocolumn,aps,pra,showpacs,superscriptaddress]{revtex4-1} 
\usepackage{graphics,dcolumn}
\usepackage{bm}
\usepackage{natbib}
\usepackage[fleqn]{amsmath}
\usepackage{hyperref}
\hypersetup{
     colorlinks=true,       
    linkcolor=blue,          
    citecolor=blue,        
    filecolor=magenta,      
    urlcolor=cyan           
}
\usepackage{exscale,relsize}
\usepackage{epsfig}
\usepackage{amssymb}
\usepackage{amsmath}
\usepackage{euscript}
\usepackage{ulem}
\usepackage{float}
\usepackage{tikz}
\usepackage{fancyvrb}
\usetikzlibrary{snakes}
\usetikzlibrary{decorations.pathmorphing}
\usetikzlibrary{patterns}
\usepackage{bbm}
\usepackage{footmisc}
 \usepackage{amsthm}

\newcommand{\be}{\begin{equation}}
\newcommand{\e}{\end{equation}}

\newcommand{\beml}{\begin{subequations}}
\newcommand{\eml}{\end{subequations}}
\newcommand{\beq}{\begin{eqnarray}}
\newcommand{\eq}{\end{eqnarray}}
\newcommand{\ba}{\begin{array}}
\newcommand{\ea}{\end{array}}

\newcommand{\lt}{\left}
\newcommand{\rt}{\right}
\newcommand{\n}{\nonumber}

\newcommand{\s}{\sigma}
\newcommand{\la}{\langle}
\newcommand{\ra}{\rangle}

\newcommand{\re}{\,{\rm Re}\,}

\begin{document}
\date{\today}

\title{Correlation functions in resonance fluorescence with spectral resolution: Signal-processing approach}

\author{Vyacheslav N. Shatokhin}
\affiliation{Physikalisches Institut, Albert-Ludwigs-Universit\"at Freiburg, Hermann-Herder-Str. 3,
D-79104 Freiburg, Germany}

\author{Sergei Ya. Kilin}
\affiliation{B. I. Stepanov Institute of Physics NASB, Nezavisimosti Avenue 68, 220072 Minsk, Belarus}


\begin{abstract} 
In the framework of the signal processing approach to single-atom resonance fluorescence with spectral resolution, we diagrammatically derive an analytical formula for arbitrary-order spectral correlation functions of the scattered fields that pass through Fabry-Perot interferometers. Our general expression is then applied to study correlation signals in the limit of well separated spectral lines of the resonance fluorescence spectrum. In particular, we study the normalized second-order temporal intensity correlation functions in the case of the interferometers tuned to the components of the spectrum and obtain interferential corrections to the approximate results derived in the secular limit. In addition, we explore purely spectral correlations and show that they can fully be understood in terms of the two-photon cascades down the dressed state ladder. 
\end{abstract}


\maketitle


\section{Introduction}
\label{sec:intro}
Single-atom resonance fluorescence (RF) has for decades served as a basic quantum electrodynamical model to study light-matter interactions  \cite{heitler,cohen_tannoudji_API}. One of the most famous features of RF is the emission spectrum of a strongly laser-driven atom consisting of three Lorentzian peaks \cite{apanasevich64,PhysRev.188.1969,schuda74,walther76,grove77} which is often referred to as the Mollow triplet. The RF spectrum provides information about the elementary scattering processes of laser photons on the atom. The identification of the scattering processes that are correlated requires a study of higher-order spectral correlation functions \cite{Apanasevich197783}.

To measure frequency resolved RF photons, one puts a spectral apparatus, such as a Fabry-Perot interferometer \cite{Eberly:77,Born_Wolf}, between a laser driven atom and a broadband detector. The presence of an interferometer in the measurement setup poses some fundamental questions, such as, what is the collapsed atomic state following detection a spectrally resolved photon\cite{Shatokhin2000157}?  Furthermore, finite resolution of any realistic filter leads to deviations of the observed (i.e., physical) RF spectra and spectral correlation functions from the ideal ones (i.e., resolved infinitely sharply).

The problem of a theoretical description of spectral detection with a proper account of the filtering process in RF from {\it real} atoms attracted much attention in the late 1970s till the early 1990s \cite{cohen-tannoudji77,apanasevich79,dalibard83,Cresser198347,nienhuis84,0022-3700-20-18-027,PhysRevA.42.503,PhysRevLett.67.2443,PhysRevA.45.8045}. In particular, temporal correlations between RF photons emitted into the components of the Mollow triplet have been studied for a wide range of filter bandwidths \cite{cohen-tannoudji77,apanasevich79,dalibard83,PhysRevLett.67.2443,PhysRevA.45.8045,0295-5075-21-3-006} in the limit of well separated spectral lines \cite{cohen-tannoudji77,apanasevich79,cohen_tannoudji_API}, providing a good quantitative agreement with the experiments \cite{PhysRevLett.45.617,PhysRevLett.67.2443,PhysRevA.45.8045}. 

In recent years, progress in coherent control of {\it artificial} atoms (e.g., quantum dots, superconducting qubits, etc.) has furnished interest in using their RF in applications, such as quantum logic devices or single-photon generators \cite{Nick-Vamivakas:2009fk}. Nonlinear optical spectroscopy with man-made quantum emitters has become a mature field where such experimental milestones as the observation of the Mollow triplet \cite{Nick-Vamivakas:2009fk,Flagg:2009,Astafiev12022010,Pigeau:2015rw,toyli16} and of the bunching of the time-ordered emission of the sideband photons \cite{UlhaqA.:2012ij} have been reached. However, the optimization of the operation regimes of devices based on artificial atoms requires a more accurate analysis of the spectral correlation functions in RF than was hitherto obtained. 
  Since RF from both real and artificial atoms can be described by essentially the same formalism, what remains is to generalize the theory of spectral correlations in RF to arbitrary driving field strengths and to arbitrary filter tuning frequencies. 

This problem has been addressed in \cite{PhysRevLett.109.183601}. Based on the approach developed in \cite{PhysRevLett.109.183601}, full two-color correlation functions of light emitted by a quantum dot have been calculated \cite{tudela13,PhysRevA.90.052111}. In particular, it has been shown \cite{PhysRevA.90.052111,PhysRevA.91.043807} that by spectrally selecting pairs of RF photons it is possible to produce frequency-entangled photon pairs or photons exhibiting strong bunching. The latter effect has recently been experimentally confirmed \cite{peiris15}.

Our present contribution is motivated not by the need to identify spectral filtration regimes that would further enhance, e.g., the nonclassical properties of RF, but rather by the following two factors.
First, within the method of \cite{PhysRevLett.109.183601}, each interferometer is treated as a separate quantum system (sensor) that is weakly coupled to a laser-driven quantum emitter. Though this approach, being in the spirit of the theory of cascaded quantum systems \cite{gardiner93,carmichael93}, is physically sound and general, its implementation demands working in the Hilbert space that is a tensor product of the constituents' Hilbert spaces. Since this implies exponentially increasing complexity with the number of sensors, it is desirable to put forward an alternative method that is free of this drawback. A method which we develop in this work represents a generalization of the so-called {\it signal-processing} approach of \cite{Cresser198347,nienhuis84,0022-3700-20-18-027,PhysRevA.42.503,PhysRevLett.67.2443,PhysRevA.45.8045}, where each filter is treated as a black box, whose output is related to the input by a spectral response function \cite{Eberly:77}. In this framework, all calculations are performed in the Hilbert space of a single atom. 

Second, in order to quantify spectral correlations, correlation functions that are normalized in one or other way have been employed in \cite{PhysRevLett.109.183601,PhysRevA.90.052111,PhysRevA.91.043807}. In the limit of well separated spectral lines, these functions attain large maxima on the tails of the spectral distribution of RF. A class of scattering processes -- the ``leapfrog'' processes -- has been introduced to explain the origin of these strong correlations \cite{tudela13}. However, it is one of the goals of this paper to show that the concept of the ``leapfrog'' processes is not justified; large values of the correlations functions stem from post-selection on the tails of the RF spectrum.

The rest of the paper is structured as follows. In the next section we recall the model of single-atom resonance fluorescence and define the general spectral correlation functions of the scattered fields that are transmitted by Fabry-Perot filters with arbitrary bandwidths and tuning frequencies. In Sec.~\ref{sec:derive} we develop our method and derive a general analytical expression for the spectral correlation functions. Next, we adapt our method to the case of temporal {\it and} spectral detection, and calculate the normalized second-order temporal intensity correlation functions of spectrally filtered fields. Section \ref{sec:3} is devoted to the application of our approach in the limit of well separated spectral lines. Thereby, we compare our results with the previous analytical results \cite{PhysRevA.45.8045} that were obtained in the {\it secular} approximation \cite{cohen_tannoudji_API}. Besides, we present an alternative explanation of the strong correlations reported in \cite{tudela13} and calculate the unnormalized spectral correlation function, which exhibits no signatures of the leapfrog transitions. We conclude our work in Sec. \ref{sec:disc}.

\section{Spectrally resolved detection of resonance fluorescence}
\label{sec:model}
We set out this section with the description of the model of single-atom resonance fluorescence. We present a master equation governing the dynamics of the atomic reduced density operator, as well as its formal solution. In Sec. \ref{sec:corr_func} we consider the problem of spectral detection of resonance fluorescence. Here we recall the relation between the normally ordered correlation functions of fields transmitted by Fabry-Perot interferometers and the multitime atomic dipole correlation functions. 

\subsection{Master equation}
Our system of interest consists of a single immobile two-level quantum emitter (an atom, molecule or a quantum dot) interacting with the quantized radiation field (bath) and with a monochromatic laser wave, whose frequency, $\omega_L$, is close to the atomic transition frequency, $\omega_{A}$: $|\omega_L-\omega_A|\ll \omega_A$. The laser field induces coherent dynamics of the atomic populations and coherences, whereas the coupling of the atom to the radiation field induces spontaneous emission as well as a decay of the off-diagonal elements of the atomic density matrix. 
The total Hamiltonian of this system reads
\be
H=H_A+H_{AL}+H_F+H_{AF},
\label{hamiltonian}
\e
where $H_A$ is a free two-level atom Hamiltonian, $H_{AL}$ is an interaction Hamiltonian of the atom with the laser field, $H_F$ is a free Hamiltonian of the radiation field, and $H_{AF}$ is an interaction Hamiltonian of the atom with the radiation field. We assume that the field bath is initially in the vacuum state and employ the standard Born-Markov and rotating wave approximations to derive a master equation governing the evolution of the reduced density matrix  of the two-level system $\rho\equiv\rho_A={\rm Tr}_F(\rho_{AF})$, averaged over the radiation field's degrees of freedom ($F$).  In the frame rotating at the laser frequency, the resulting master equation reads \cite{carmichael}:
\begin{align}
\dot{\rho}&={\cal L}\rho=i\frac{\Delta}{2}[\sigma_z,\rho]-i\frac{v}{2}[\sigma_++\sigma_-,\rho]\n\\
&+\gamma(2\sigma_-\rho\sigma_+-\sigma_+\sigma_-\rho-\rho\sigma_+\sigma_-),
\label{meq}
\end{align}  
where ${\cal L}$ is a Liouvillian superoperator , $\sigma_+=|2\ra\la1|$, $\sigma_-=|1\ra\la2|$ and $\sigma_z=|2\ra\la2|-|1\ra\la1|$ are respectively the atomic raising, lowering and inversion operators, $\Delta=\omega_L-\omega_A$, $v$ is the Rabi frequency, and $\gamma$ is half the spontaneous decay rate.

Equation (\ref{meq}) has a formal solution, 
\be
\rho(t)=e^{{\cal L}(t-t_0)}\rho(t_0),\n
\e
or, in a matrix form,
\be
\rho_{kl}(t)=\sum_{i,j=1}^{2}{\cal D}_{kl}^{ij}(t-t_0)\rho_{ij}(t_0), 
\label{Green}
\e
where ${\cal D}_{kl}^{ij}(t)$ are the matrix elements of the Green's matrix  \cite{apan78} of Eq. (\ref{meq}). By virtue of the quantum regression theorem \cite{scully}, arbitrary multitime dipole correlation functions for an atom whose dynamics is governed by the Markov master equation of type (\ref{meq}) can be expressed through products of ${\cal D}_{kl}^{ij}(t)$ \cite{apan78} (see Sec. \ref{sec:diag}).

\subsection{Correlation functions of spectrally filtered fields}
\label{sec:corr_func}
According to Glauber's photodetection theory \cite{glauber}, quantum statistical properties of the electromagnetic field can be characterized by a set of the normally ordered field correlation functions, \begin{align}
&G^{(n,m)}({\bf x}_1,\ldots, {\bf x}_n, {\bf x}_{n+1}, \ldots ,{\bf x}_{n+m})\n\\
&=\!\la {\bf E}^{(-)}({\bf x}_1)\!\ldots \!{\bf E}^{(-)}({\bf x}_n) {\bf E}^{(+)}({\bf x}_{n+1})\!\ldots\!{\bf E}^{(+)}({\bf x}_{n+m})\ra,
\label{corr_mn}
\end{align}  
where ${\bf E}^{(+/-)}({\bf x}_i)$ denotes the positive-/negative-frequency part of the electric field vector operator at the space-time point ${\bf x}_i\equiv \{{\bf r}_i,t_i\}$. 

Let us consider spectral detection of single-atom resonance fluorescence, whereupon each of the field components scattered by the atom is spectrally resolved by an interference filter. Then the positive-frequency component of the field can be represented as the following sum \cite{Knoll:86,vogel}:
\be
{\bf E}^{(+)}({\bf x}_i)={\bf E}_{\rm free}^{(+)}({\bf x}_i)+{\bf E}_{\rm s}^{(+)}({\bf x}_i),
\label{E_pos}
\e
where the first and second terms in the right-hand side are, respectively, the free- and source-field components. Due to the vacuum initial state of the radiation field, the free-field component does not contribute to the normally ordered averages of the field operators, and will be dropped in subsequent expressions. 

The spectrally resolved field of the atomic source is given by the convolution \cite{vogel,Eberly:77}
\be
{\bf E}_{\rm s}^{(+)}({\bf r},t+\Delta t+|{\bf r}|/c)\propto \int_0^{t}dt^\prime T_{\rm f}(t-t^\prime)\sigma_-(t^\prime),
\label{filtered_E}
\e
where $T_{\rm f}(t)$ is the filter transmission function, $\Delta t$ is the time delay caused by difference between the speed of light in a dielectric medium of the filter and in vacuum, and $\sigma_-(t)$ is the atomic lowering operator in the frame rotating at the laser frequency. In passing, we note that the integral form (\ref{filtered_E}) is typical of non-Markov processes \cite{breuer_book,PhysRevA.59.2306}, since the filtered field at time $t$ is determined by the distribution of the atomic emission events over the entire atom-laser field interaction history.  

Throughout this work, we assume our filtering devices to be Fabry-Perot interferometers, whose transmission response functions can be approximated by a single exponential \cite{Eberly:77},
\begin{align}
T_{\rm f}(t)&=\Theta(t)\Gamma e^{-(\Gamma+i\delta)t},\n\\
&\equiv\Theta(t){\rm Re}[\lambda]e^{-\lambda t},
\label{trans_f}
\end{align}
where $\Theta(t)$ is the unit step function, $\Gamma$ is the filter bandwidth, and $\delta=\omega-\omega_L$ is the detuning between the filter and laser frequencies. 

Before we move on, we would like to mention a recent work on spectral correlations of photons emitted by a laser driven single molecule \cite{PhysRevLett.102.018303}, where the filtering device has implicitly  been referred to as the Fabry-Perot interferometer. However,  
the expression for the frequency resolved field correlation function in \cite{PhysRevLett.102.018303} differs from that given by Eq. (\ref{G^n}) and rather corresponds to the spectral decomposition performed by a {\it prism} \cite{PhysRevA.59.2306}.

Equation  (\ref{filtered_E}) can be simplified if we ignore the retardation effects and set $\Delta t+|{\bf r}|/c=0$ for each atom-filter-detector path. This approximation becomes exact in the steady state limit $t\to\infty$, on which we will focus henceforth. Furthermore, we assume equal optical paths from the atom to each detector. In this case the spatial dependence in (\ref{filtered_E}) can be dropped and we arrive at the following expression for the correlation function of the spectrally resolved fields \cite{0022-3700-20-18-027,PhysRevA.42.503}
\begin{widetext}
\begin{align}
G&^{(n,m)}(\lambda_1,\ldots, \lambda_n,\lambda_{n+1}, \ldots, \lambda_{n+m})
=\lim_{t\to\infty}\int_0^tdt_1\ldots\int_0^tdt_n\int_0^tdt_{n+1}\ldots\int_0^tdt_{n+m}\n\\
&\times T_{\rm f_1}^*(t-t_1)\ldots T^*_{\rm f_n}(t-t_n)T_{\rm f_{n+1}}(t-t_{n+1})\ldots T_{\rm f_{n+m}}(t-t_{n+m})
\lt\la\overrightarrow{T}\lt[\sigma_+(t_1)\ldots\sigma_+(t_n)\rt]\overleftarrow{T}\lt[\sigma_-(t_{n+1})\ldots\sigma_-(t_{n+m})\rt]\rt\ra,
\label{G^n}
\end{align}
\end{widetext}
where $\overrightarrow{T}[\ldots]$ ($\overleftarrow{T}[\ldots]$) are the operators of chronological ordering which arrange the atomic raising (lowering) operators such that their time arguments increase from left to right (from right to left), as indicated by the arrows. Expression (\ref{G^n}) corresponds to simultaneous detection of $n+m$ field components (since $t_i=t$ for $i=1,\ldots,n+m$) in a setup where each positive- and negative-frequency field component is filtered with an individual filter. 

In the particular case of spectrally resolved {\it intensity} correlation functions, one sets $m=n$ and $T_{{\rm f}_i}(t)=T_{{\rm f}_{2n+1-i}}(t)$ ($i=1,\ldots, n$). 
After we present a recipe for calculating the functions $G^{(n,m)}(\lambda_1,\ldots, \lambda_n,\lambda_{n+1}, \ldots, \lambda_{n+m})$ in Sec. \ref{sec:derive},  we will focus on spectral {\it and} temporal detection of resonance fluorescence for $n=m=2$ in Sec. \ref{sec:int_tau}.  In the latter case, the intensity correlation function depends not only on the parameters of two spectrometers, but also on a time delay, $\tau$, between the detection events of spectrally filtered photons. 

\section{Derivation of a general formula for $G^{(n,m)}(\lambda_1,\ldots, \lambda_{n+m})$}
\label{sec:derive}
In this section, we take the multifold integrals in the right hand side of Eq. (\ref{G^n}). We divide this task into two steps. In Sec. \ref{sec:diag}, we introduce diagrams that allow us to express the multitime dipole correlation function in a transparent way. In Sec. \ref{sec:integrals}, we use the Laplace transform to obtain the analytical expression for this function.

\subsection{Diagrammatic presentation of the multitime dipole correlation functions}
\label{sec:diag}
As seen from Eq. (\ref{G^n}), the spectral field correlation function represents a multifold convolution of the atomic dipole correlation function, 
\begin{align}
C^{(n,m)}&(t_1,\ldots,t_{n+m})\equiv \la\overrightarrow{T}[\sigma_+(t_1)\ldots\sigma_+(t_n)]\n\\
&\times\overleftarrow{T}[\sigma_-(t_{n+1})\ldots\sigma_-(t_{n+m})]\ra,
\end{align}
with the filters' transmission functions. The calculation of  $C^{(n,m)}(t_1,\ldots,t_{n+m})$ is complicated by the fact that the atomic operators do not commute with themselves at different times \cite{vogel}. Therefore, in order to find $C^{(n,m)}(t_1,\ldots,t_{n+m})$, one needs to split the multiple integral in the right hand side of (\ref{G^n}) into a sum of $(n+m)!$ time-ordered integrals and apply the quantum regression theorem to each of the resulting functions $C^{(n,m)}(t_1,\ldots,t_{n+m})$, whose arguments  now have a definite order. 

In the following we show that this task -- the calculation of the convolution integrals of the time-ordered correlation functions $C^{(n,m)}(t_1,\ldots,t_{n+m})$ -- can be accomplished, and a general analytical expression for the function $G^{(n,m)}(\lambda_1,\ldots, \lambda_{n+m})$ can be derived for arbitrary $n,m$.

\begin{center}
\begin{figure}
{\includegraphics[width=7cm]{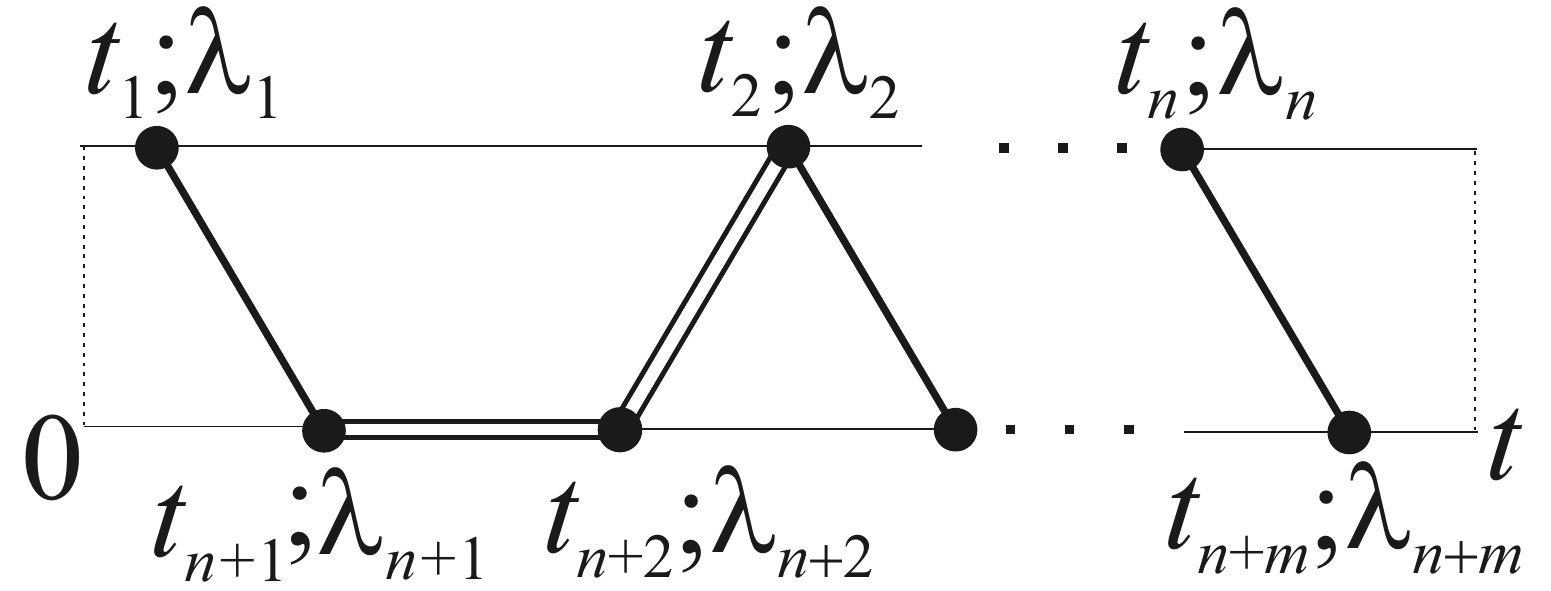}}
\caption{
Example of a diagram that corresponds to the following time order: 
$0\leq t_1\leq t_{n+1}\leq t_{n+2}\leq t_2\leq t_{n+3}\leq \ldots\leq t_{n}\leq t_{n+m}\leq t$. Beside each time $t_i$ we include the exponents $\lambda_i$ of the filter transmission functions (\ref{trans_f}), with which the operators $\sigma_\pm(t_i)$ are convolved. Subsequent times are connected by single or double lines that are associated with the propagators ${\bf D}^{[+]}(t)$ and ${\bf D}^{[-]}(t)$, respectively [see Eq. (\ref{DDr})].}
\label{fig0}
\end{figure}
\end{center}

 According to the definition of the normally ordered correlation function (\ref{G^n}), times $t_1,\ldots, t_n$ are associated with the atomic raising operator $\sigma_+$, whereas times $t_{n+1},\ldots, t_{n+m}$ are associated with the atomic lowering operator $\sigma_-$. It is instructive to represent an arbitrary temporal sequence using double-row diagrams, where the upper and lower rows carry times associated with the operators $\sigma_+$ and $\sigma_-$, respectively. Figure \ref{fig0} gives an example of a possible order of times $t_1,\ldots, t_n$,  namely, $0\leq t_1\leq t_{n+1}\leq t_{n+2}\leq t_2\leq t_{n+3}\leq \ldots\leq t_{n}\leq t_{n+m}\leq t$. This type of diagram is somewhat reminiscent of the double-sided Feynman diagrams that have been extensively used in nonlinear optical spectroscopy \cite{mukamel_book}.
Subsequent times are connected by a single or a double line (see Fig. \ref{fig0}) in accordance with the following rule: If the line's outgoing time is in the upper row then the line is single; otherwise, it is double. These lines correspond to two types of matrix propagators that are needed to assess the multitime dipole correlation functions. As already mentioned, such correlation functions can be 
expressed through products of the Green's matrix elements ${\cal D}^{ij}_{kl}(t)$ [$i,j,k,l=1,2$; see Eq. (\ref{Green})], whose total number is 16. However, since the function $C^{(n,m)}(t_1,\ldots,t_{n+m})$ includes two types of atomic operators ($\sigma_+$ and $\sigma_-$), only nine elements of  ${\cal D}^{ij}_{kl}(t)$ suffice to calculate 
$C^{(n,m)}(t_1,\ldots,t_{n+m})$. Consistently, three out of four density matrix elements $\rho_{ij}(t_0)$ come into play; they can be arranged into a three-component vector. The resulting matrix propagators and vector read
\beml
\begin{align}
{\bf D}^{[+]}(t)&\!=\!\lt(\begin{array}{ccc}{\cal D}^{11}_{12}(t)&0&{\cal D}^{21}_{12}(t)\\
{\cal D}^{11}_{21}(t)&0&{\cal D}^{21}_{21}(t)\\
{\cal D}^{11}_{22}(t)&0&{\cal D}^{21}_{22}(t)
\end{array}\rt), \\
{\bf D}^{[-]}(t)&\!=\!\lt(\begin{array}{ccc}0&{\cal D}^{11}_{12}(t)&{\cal D}^{12}_{12}(t)\\
0&{\cal D}^{11}_{21}(t)&{\cal D}^{12}_{21}(t)\\
0&{\cal D}^{11}_{22}(t)&{\cal D}^{12}_{22}(t)
\end{array}\rt), \label{Dpm}\\
{\bf r}(t)&=\lt[\rho_{12}(t),\rho_{21}(t),\rho_{22}(t)\rt]^T,
\label{rt}
\end{align}
\label{DDr}
\eml
where ${\bf D}^{[+]}(t)$ and ${\bf D}^{[-]}(t)$ correspond to single and double lines in a diagram, respectively.  
The computation of $C^{(n,m)}(t_1,\ldots,t_{n+m})$ now reduces to the multiplication of the vector ${\bf r}(t)$, taken at the earliest time, by the propagators between subsequent times. Finally, $C^{(n,m)}(t_1,\ldots,t_{n+m})$ is given by the first (second) element of the resulting vector, if the final time is in the upper (lower) row. As will become clear shortly, it is convenient to denote these first and second elements as $\{.\}_+$ and $\{.\}_-$, respectively. 

The above rules provide an unambiguous way to find $C^{(n,m)}(t_1,\ldots,t_{n+m})$. For example, the expression for the multi-time correlation function represented by the diagram in Fig. \ref{fig0} reads
\begin{align}
&C^{(n,m)}(t_1,\!\ldots\!,t_{n+m})\!=\!\{{\bf D}^{[+]}(t_{n+m}\!-\!t_n)\!\ldots\! \n\\
&\times{\bf D}^{[+]}(t_{n+3}\!-\!t_2){\bf D}^{[-]}(t_2\!-\!t_{n+2}){\bf D}^{[-]}(t_{n+2}\!-\!t_{n+1})\n\\
&\times{\bf D}^{[+]}(t_{n+1}\!-\!t_1){\bf r}(t_1)\}_+,
\label{time_corr}
\end{align}
and its generalization to an arbitrary double-row diagram (i.e., arbitrary time ordering) is straightforward. 

\subsection{Calculation of the multifold convolution integrals}
\label{sec:integrals}
Now, expanding the right-hand side of (\ref{G^n}) into a sum of $(n+m)!$ time ordered integrals, and using Eqs. (\ref{trans_f}) and (\ref{time_corr}), we arrive at the following expression for the function $G^{(n,m)}$ (for brevity, we omit its arguments):
\begin{align}
G^{(n,m)}&\!=
\!\!\!\!\!\!\!\!\sum_{\pi(j_1,\ldots,j_{n+m})}\!\!\!\!\!\!\!\lim_{t\to\infty}\!\int_0^t\!\!dt_{j_1}\int_0^{t_{j_1}}\!\!\!dt_{j_2}\ldots \!\!\!\int_0^{t_{j_{n+m-1}}}\!\!dt_{j_{n+m}} \n\\
&\times\!\!\! \prod_{k=1}^{n+m}\!\!\Gamma_k e^{-\lambda_{j_k}(t-t_{j_k})}\{{\bf D}^{[s_{j_2}]}(t_{j_1}\!-\!t_{j_2})\!\ldots\n\\
&\times \! {\bf D}^{[s_{j_{n+m}}]}(t_{j_{n+m-1}}\!-\!t_{j_{n+m}}){\bf r}(t_{j_{n+m}})\}_{s_{j_1}},
\label{int_t}
\end{align}
where $\pi(j_1,\ldots,j_{n+m})$ denotes permutations of indices $j_1,\ldots,j_{n+m}\in\{1,\ldots,n+m\}$, and
\be
s_{j_k}=
\begin{cases}
+ &\text{if $j_k\in\{1,\ldots,n\}$,}\\
-  &\text{if $j_k\in\{n+1,\ldots,n+m\}$.}
\end{cases}
\e

Due to the exponential form of the filter transmission functions (\ref{trans_f}), the latter are nothing but kernels of the Laplace transforms, with variables $\lambda_k$ (note that for each $k$, $\re[\lambda_k]=\Gamma_k>0$), shown beside the respective times, $t_k$, in Fig. \ref{fig0}. This allows us to take the convolution integrals in Eq. (\ref{int_t}) exactly, with the result (see Appendix \ref{conv1})
\be
G^{(n,m)}
\!=\!\frac{1}{\Lambda_1}\!\!\sum_{\pi(j_1,\ldots,j_{n+m})}\!\!\!\Gamma_{j_1}\lt\{\lt[\prod_{k=2}^{n+m}\Gamma_{j_k}\tilde{\bf D}^{[s_{j_k}]}(\Lambda_k)\rt]{\bf r}_\infty\rt\}_{s_{j_1}},
\label{aux0}
\e
where  ${\bf r}_\infty=\lim_{t\to\infty}{\bf r}(t)$, 
\be
\Lambda_k=\sum_{l=k}^{n+m}\lambda_{j_l},
\e
and $\tilde{\bf D}^{[\pm]}(p)$ is Laplace transform of the propagator ${\bf D}^{[\pm]}(t)$. The explicit expressions for the elements of $\tilde{\bf D}^{[\pm]}(p)$ and ${\bf r}_\infty$ are given in Appendix \ref{green_a}.
Thus, according to our result (\ref{aux0}), the calculation of the stationary spectrally resolved correlation function $G^{(n,m)}$ amounts to a sum of $(n+m)!$ products of $(n+m-1)$ matrices $\tilde{\bf D}^{[\pm]}(p)$ [see Eq. (\ref{Dpm})]; a task which can be efficiently implemented numerically. 

It should be noted that the structure of the expression (\ref{aux0}) is similar to that of single-atom spectral response functions that appear in the multiple scattering theory of intense laser light from cold atoms \cite{shatokhin12b}. The precise relationship between these two types of functions will be established in future work. 

For the particular case of spectrally resolved {\it intensity} correlation functions, we have $m=n$ and $\lambda_k=(\lambda_{2n+1-k})^*$ ($k=1,\ldots, n$). The correlation function $G^{(n,n)}$ then depends on $n$ bandwidths and $n$ detunings (instead of $2n$ bandwidths and  $2n$ detunings): $G^{(n,n)}\equiv G^{(n,n)}(\Gamma_1,\delta_1;\ldots;\Gamma_n,\delta_n)$. 

For $n=m=1$, Eq. (\ref{aux0}) reduces to the first-order spectral field correlation function, which is related to the stationary physical spectrum, $S(\Gamma,\delta)$, via $S(\Gamma,\delta)=\Gamma^{-1}G^{(1,1)}(\Gamma,\delta)$  \cite{vogel}, or
\be
S(\Gamma,\delta)=\re\lt[\lt\{\tilde{\bf D}^{[+]}(\Gamma-i\delta){\bf r}_\infty\rt\}_-\rt].
\label{def_G11}
\e 
\begin{center}
\begin{figure}
{\includegraphics[width=7cm]{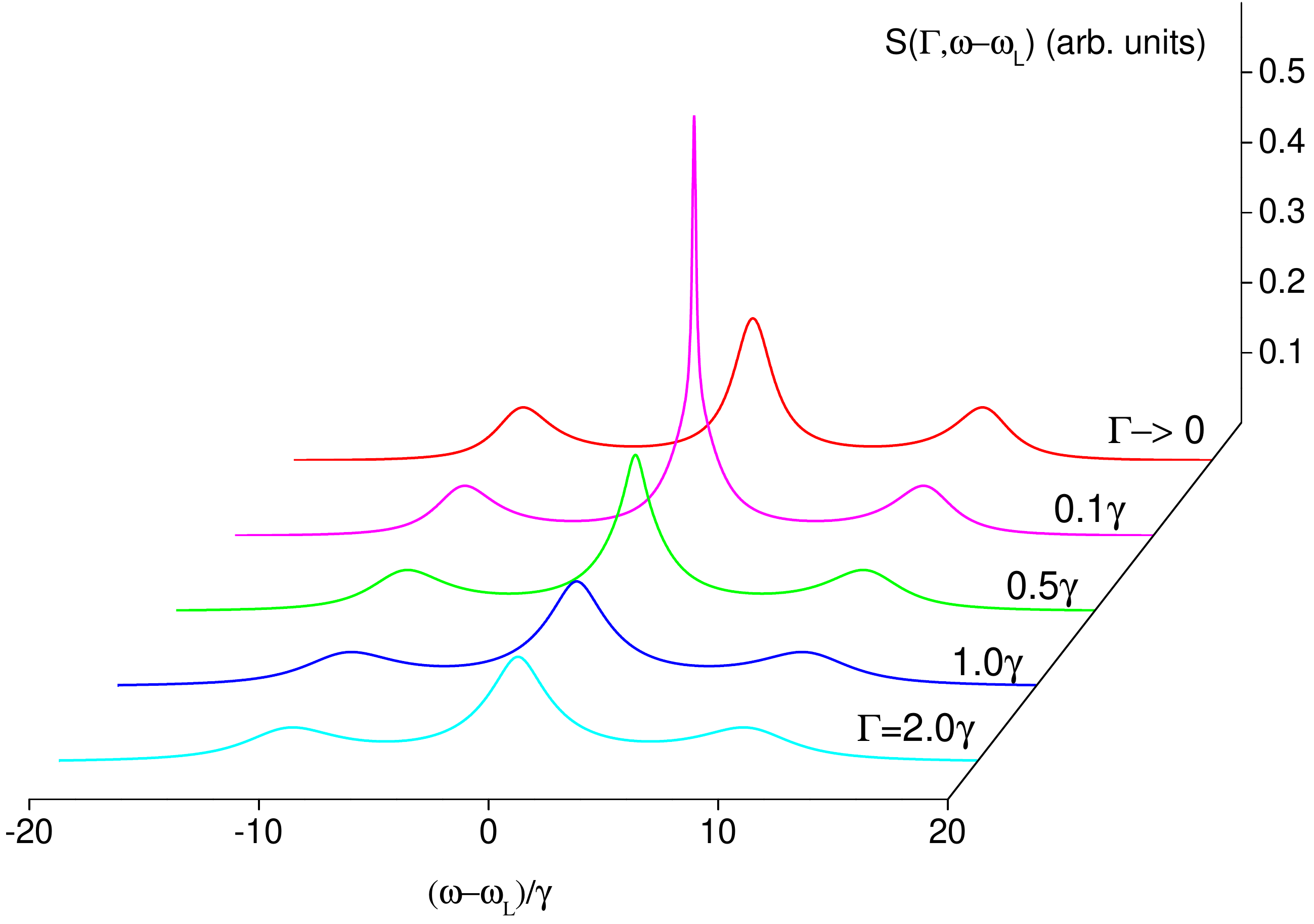}}
\caption{
(Color online) Stationary physical spectra $S(\Gamma,\omega-\omega_L)$, calculated using Eq. (\ref{def_G11}) at $v=10\gamma$ and $\Delta=2\gamma$ (i.e., $\Omega\approx 10.2\gamma$) for filter bandwidths $\Gamma/\gamma=0,0.1,0.5,1.0,2.0$. The Mollow triplet corresponds to $\lim_{\Gamma\to 0}S(\Gamma,\omega-\omega_L)$.}
\label{spectra}
\end{figure}
\end{center}

We illustrate the $\Gamma$ dependence of the physical spectra [see Eq. (\ref{def_G11})] at a moderate generalized Rabi frequency, $\Omega\approx 10.2\gamma$, by several examples in Fig. \ref{spectra}. Like the Mollow triplet, the stationary physical spectrum consists of three Lorentzian components that are symmetrically shifted by $\Omega$ from the central peak, located at the laser frequency. Although the line shape of the physical spectra, in general, deviates from that of the Mollow triplet \cite{PhysRev.188.1969}, it tends to the latter in the limit $\Gamma\to 0$.

Let us finally note that the expression (\ref{def_G11}) for the stationary physical spectrum can be generalized to a multilevel atom, where it is possible to control interference between emission processes from different dipole transitions of the atom through variation of the filter bandwidth \cite{PhysRevLett.96.100403,PhysRevA.73.063814}.

\section{Temporal and spectrally resolved detection. Case $n=m=2$}
\label{sec:int_tau}
Having obtained the general expression, Eq. (\ref{aux0}), for the function $G^{(n,m)}$, we now specialize on the case of $n=m=2$, where the two photodetection events are separated by a time delay $\tau$. From the definition of the temporal intensity correlation function, we move on to a discussion of its symmetry properties with respect to the filter tuning frequencies at $\tau=0$.  Finally, we consider arbitrary time delays and introduce diagrams that help us to take the emerging convolution integrals. 

\subsection{Definition}
\label{sec:def}
In this section, we consider the case of simultaneous temporal and spectral detection, focusing on the second-order temporal intensity correlation function of the spectrally resolved fields. 
For equal filter bandwidths, $\Gamma_1=\Gamma_2=\Gamma$, this function is defined as
\begin{align}
G^{(2,2)}_\tau&(\Gamma;\delta_1,\delta_2)
\!=\!\lim_{t\to\infty}\!\int_0^t\!dt_1\int_0^{t+\tau}\!dt_2\!\int_0^{t+\tau}\!dt_3\int_0^t\!dt_4\n\\
&\times \!T_{\rm f_1}^*(t\!-\!t_1)T^*_{\rm f_2}(t\!-\!t_2)T_{\rm f_2}(t\!-\!t_3)T_{\rm f_1}(t\!-\!t_4)\n\\
&\times \!\lt\la\overrightarrow{T}\!\lt[\sigma_+(t_1)\sigma_+(t_2)\rt]\overleftarrow{T}\!\lt[\sigma_-(t_{3})\sigma_-(t_{4})\rt]\rt\ra,
\label{G22}
\end{align}
where time delay $\tau\geq 0$ corresponds to the case where a photon resolved by the interferometer with the detuning $\delta_1$ is detected first. 

\subsection{Symmetry properties at $\tau=0$}
\label{sec:symmetry}
The zero-delay second-order intensity correlation function reads [compare to Eq.~(\ref{aux0})],
\be
G^{(2,2)}_0(\Gamma,\delta_1,\delta_2)
\!\equiv \!\frac{\Gamma^3}{4}\!\sum_{\pi(j_1,\ldots,j_{4})}\!\lt\{\lt[\prod_{k=2}^{4}\tilde{\bf D}^{[s_{j_k}]}\lt(\!\sum_{l=k}^{4}\lambda_{j_l}\!\rt)\!\rt]{\bf r}_\infty\!\rt\}_{s_{j_1}}\!\!\!.
\label{aux220}
\e
Henceforth, all spectral correlation functions corresponding to simultaneous detection ($\tau=0$) will for definiteness be furnished by the subscript 0. Consistently, the notation $G^{(1,1)}_0$ will be reserved for the stationary first-order field correlation function. 

For simultaneous detection, the order of the detunings $\delta_1$ and $\delta_2$  in Eq.~(\ref{aux220}) becomes immaterial, which leads to the mirror reflection symmetry about the diagonal in the ($\delta_1$, $\delta_2$) plane,
\be
G^{(2,2)}_0(\Gamma,\delta_1,\delta_2)=G^{(2,2)}_0(\Gamma,\delta_2,\delta_1).
\label{sym_diag}
\e
Furthermore, at exact resonance ($\Delta=0$), there appears an additional mirror reflection symmetry about the antidiagonal $\delta_1=-\delta_2$: 
\be
G^{(2,2)}_0(\Gamma,\delta_1,\delta_2)=G^{(2,2)}_0(\Gamma,-\delta_1,-\delta_2).
\label{sym_antidiag}
\e
Both symmetry properties can be explicitly demonstrated by the perturbative calculation of the correlation signals using the two-photon scattering amplitudes, which is valid in the limit $v\ll \gamma$. From Eq. (\ref{aux220}) we obtain the result
\be
G^{(2,2)}_0=\lt(\frac{\Gamma v}{2}\rt)^4\frac{P}{Q}+O\lt((v/\gamma)^5\rt),
\e
where
\beml
\begin{align}
P&=\!8\Gamma \gamma\delta_1\delta_2\!+8\gamma\Gamma^3\!+\!4\Gamma^2[\Delta^2\!+\!2\delta_1\delta_2\!-\!\Delta(\delta_1\!+\!\delta_2)]\n\\
&\!+\!4\Gamma^4\!+\![\delta_1^2\!+\!\delta_2^2\!-\!\Delta(\delta_1\!+\!\delta_2)]^2\!+\!\gamma^2[4\Gamma^2\!+\!(\delta_1\!+\!\delta_2)^2],\\
Q&=\!(\gamma^2\!+\!\Delta^2)(\Gamma^2\!+\!\delta_1^2)(\Gamma^2\!+\!\delta_2^2)[4\Gamma^2\!+\!(\delta_1\!+\!\delta_2)^2]\n\\
&\times[(\Gamma\!+\!\gamma)^2\!+\!(\Delta\!-\!\delta_1)^2][(\Gamma\!+\!\gamma)^2\!+\!(\Delta\!-\!\delta_2)^2].
\end{align}
\eml
In the nonperturbative regime, Eqs. (\ref{sym_diag}) and (\ref{sym_antidiag}) also hold true. In this case the symmetry $G^{(2,2)}_0(\Gamma,\delta_1,\delta_2)=G^{(2,2)}_0(\Gamma,-\delta_1,-\delta_2)$, at $\Delta=0$, is related to the equal state populations of the eigenstates of laser-atom interaction Hamiltonian ({\it dressed} states \cite{cohen_tannoudji_API}; see Sec. \ref{secular}).
 
\subsection{Decomposition of Eq.~(\ref{G22})}
Without loss of generality, we assume that $\tau\geq 0$ (the result for $\tau<0$ follows upon the replacements $\delta_1\leftrightarrow \delta_2$), and expand the fourfold integral in Eq. (\ref{G22}) into four terms:
\be
G^{(2,2)}_\tau=\sum_{k=1}^4I_k(\tau),
\label{def_G22tau}
\e
where
\beml
\begin{align}
I_1(\tau)&=\lim_{t\to\infty}\int_0^tdt_1\int_0^{t}dt_2\int_0^{t}dt_3\int_0^tdt_4\n\\
&\times F_{(t,\tau)}(t_1\ldots,t_4),\label{i1}\\
I_2(\tau)&=\lim_{t\to\infty}\int_0^tdt_1\int_t^{t+\tau}dt_2
\int_0^{t}dt_3\int_0^tdt_4\n\\
&\times F_{(t,\tau)}(t_1\ldots,t_4),\label{i2}\\
I_3(\tau)&=\lim_{t\to\infty}\int_0^tdt_1\int_0^{t}dt_2
\int_t^{t+\tau}dt_3\int_0^tdt_4\n\\
&\times F_{(t,\tau)}(t_1\ldots,t_4),\label{i3}\\
I_4(\tau)&=\lim_{t\to\infty}\int_0^tdt_1\int_t^{t+\tau}dt_2
\int_t^{t+\tau}dt_3\int_0^tdt_4\n\\
&\times F_{(t,\tau)}(t_1\ldots,t_4),\label{i4}
\end{align}
\eml
with
\begin{align}
F_{(t,\tau)}&(t_1\ldots,t_4)=\Gamma^4e^{-2\Gamma\tau}\prod_{k=1}^4e^{-\lambda_k(t-t_k)}\n\\
&\times\!\lt\la\overrightarrow{T}\!\lt[\sigma_+(t_1)\sigma_+(t_2)\rt]\!\overleftarrow{T}\!\lt[\sigma_-(t_{3})\sigma_-(t_{4})\rt]\rt\ra.
\label{def_F}
\end{align}
Since $I_1(\tau)$ coincides, up to the exponential prefactor $e^{-2\Gamma\tau}$, with $G^{(2,2)}_0$ [see Eqs. (\ref{i1}), (\ref{def_F}), and (\ref{G22})], we obtain 
\be
I_1(\tau)=e^{-2\Gamma\tau}G^{(2,2)}_0,
\e
where $G^{(2,2)}_0$ is given by Eq. (\ref{aux220}).

\begin{center}
\begin{figure}
{\includegraphics[width=8cm]{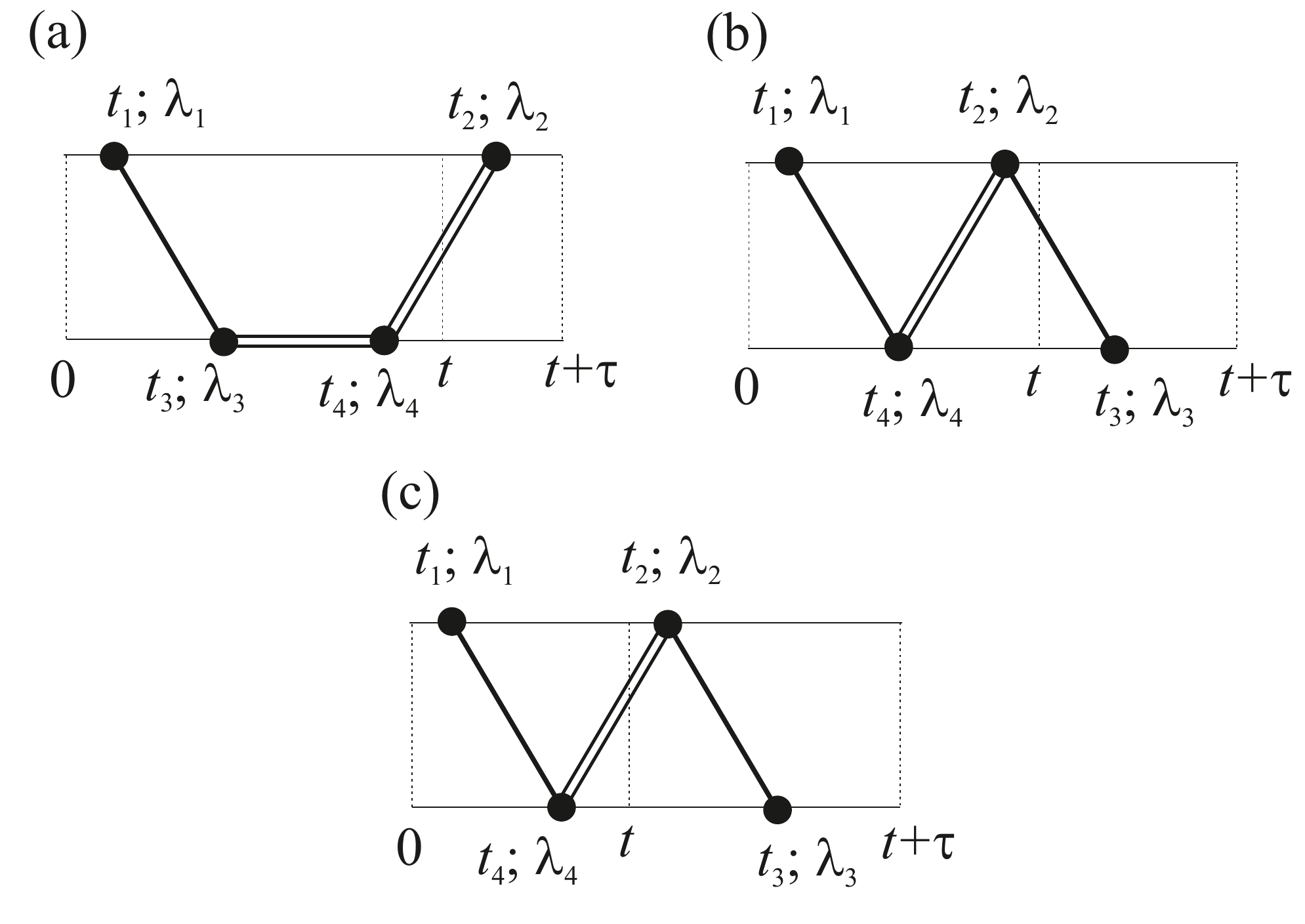}}
\caption{
Examples of time-ordered diagrams that contribute to (a) $I_2(\tau)$, (b) $I_3(\tau)$, and (c) $I_4(\tau)$.}
\label{fig2}
\end{figure}
\end{center}

The remaining three integrals, given by Eqs. (\ref{i2})-(\ref{i4}), are partially temporarily ordered.
 By analogy with the case of the function $G^{(n,m)}$ (see Sec. \ref{sec:diag}), we expand these integrals into fully time-ordered ones, and use double-row diagrams to calculate the dipole correlation functions  $\lt\la\overrightarrow{T}\lt[\sigma_+(t_1)\sigma_+(t_2)\rt]\overleftarrow{T}\lt[\sigma_-(t_{3})\sigma_-(t_{4})\rt]\rt\ra$. Examples of fully time-ordered diagrams corresponding to integrals $I_2(\tau)$, $I_3(\tau)$, and $I_4(\tau)$ are presented in Fig. \ref{fig2}. 
It is easy to see that there are overall six terms in the expansions of $I_2(\tau)$ and $I_3(\tau)$ into the fully time-ordered integrals. As for $I_4(\tau)$, it can be decomposed into four fully time-ordered integrals. The derivation of the analytical expressions for $I_2(\tau)$, $I_3(\tau)$, and $I_4(\tau)$ is given in Appendix \ref{time_delay}. 

Using the results obtained in this section, we will next analyze the behavior of the second-order temporal intensity correlation function of frequency resolved RF in the limit of well separated spectral lines.    

\section{Numerical results: Limit of well-separated components}
\label{sec:3}
In this section, we apply the general expressions that we derived in Secs. \ref{sec:derive} and \ref{sec:int_tau} to analyze correlation signals in resonance fluorescence with spectral resolution in the limit of well-separated components,  
$\Omega\gg\gamma$. Thereby, we attain two goals. One the one hand, we obtain corrections to the approximate expressions for the second-order temporal intensity correlation functions that were obtained in \cite{PhysRevA.45.8045}. On the other hand, it is in this regime of strong atom-laser field coupling where the authors of \cite{tudela13} introduced a new class of elementary processes to explain the behavior of the normalized second-order intensity correlation functions with spectral resolution. We show that the observed features stem from normalization, whereas spectral correlations can be understood using the ``standard'' transitions down the dressed-state ladder.

\subsection{Normalized second-order intensity correlation function}
\label{well-sep}
\subsubsection{Approximate versus rigorous treatment}
\label{secular}
\begin{center}
\begin{figure}
{\includegraphics[width=5cm]{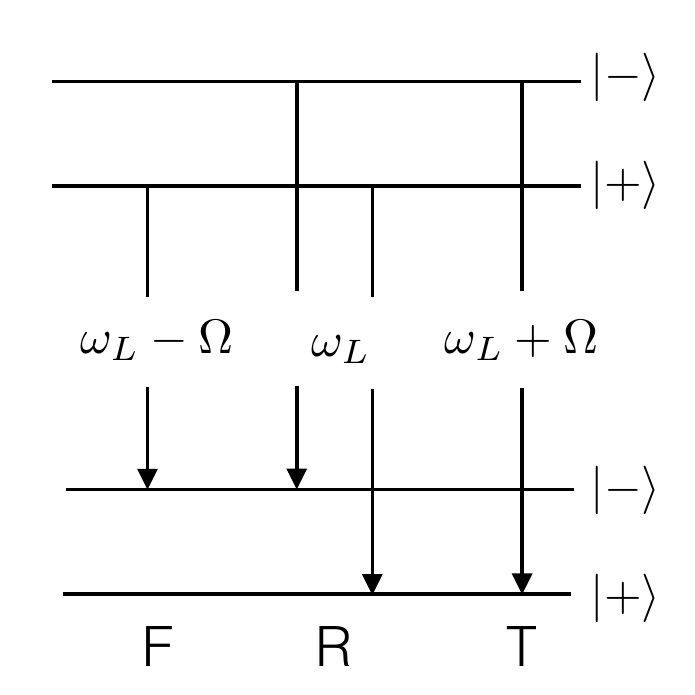}}
\caption{
Dressed levels, $|+\ra$ and $|-\ra$ [see Eq. (\ref{def_dressed})], and spontaneous transitions giving rise the resonance fluorescence triplet in the limit of well-separated spectral lines. The components of the triplet are centered at the frequencies $\omega_L-\Omega$, $\omega_L$, and $\omega_L+\Omega$, which for $\Delta=\omega_L-\omega_0>0$ are referred to as fluorescence (F), Rayleigh (R), and three-photon (T) lines, respectively. }
\label{dressed_levels}
\end{figure}
\end{center}

As already mentioned, in the limit $\Omega\gg\gamma$, where $\Omega=(\Delta^2+v^2)^{1/2}$ is the generalized Rabi frequency, the emission spectrum of resonance fluorescence splits into three components \cite{PhysRev.188.1969}, each of which has a width of the order of $\gamma$, that are centered at the well-separated frequencies $\omega_L-\Omega$, $\omega_L$, and $\omega_L+\Omega$. In this case, the spectral lines of the RF triplet can be attributed to spontaneous transitions down the ladder of the so-called {\it dressed} states \cite{cohen-tannoudji77} (see Fig. \ref{dressed_levels}), 
\beml
\begin{align}
|-\ra&=c_{\theta/2}|1\ra-s_{\theta/2}|2\ra,\\
|+\ra&=s_{\theta/2}|1\ra+c_{\theta/2}|2\ra,
\end{align}
\label{def_dressed}
\eml
where $\theta=\arccos(\Delta/\Omega)$, and $c_x\equiv \cos x$, $s_x\equiv \sin x$. States $|\pm\ra$
are the eigenstates of the laser-atom interaction Hamiltonian [upper line of Eq. (\ref{meq})]. An analysis of the transitions between the dressed states makes it possible not only to interpret the RF triplet, but also to identify temporal correlations between the components thereof  \cite{apanasevich79,Cohen-Tannoudji79}. However, the results of \cite{apanasevich79,Cohen-Tannoudji79} do not explicitly include spectral filters, which can alter the statistics of the detected photons \cite{Armstrong:66}. Furthermore, the treatments of \cite{apanasevich79,Cohen-Tannoudji79} are based on the {\it secular} approximation (see below). The description of photon correlations between the components of the Mollow triplet in the secular limit, but with the incorporation of the frequency filters, has been done in \cite{PhysRevLett.67.2443,PhysRevA.45.8045,0295-5075-21-3-006,1464-4266-2-2-317}. Some of the theoretical predictions \cite{PhysRevLett.67.2443,PhysRevA.45.8045} have found a good agreement with the experimental observation using broad filters, $\gamma\ll\Gamma\ll\Omega$ \cite{PhysRevLett.67.2443,PhysRevA.45.8045}. 

In this section, we present the results of our rigorous calculations of the temporal correlations between the peaks of the RF triplet. Thereby, we obtain corrections to the previous approximate results \cite{PhysRevLett.67.2443,PhysRevA.45.8045} which are consistent with the small error due to the secular approximation \cite{cohen_tannoudji_API}.
We recall that, to derive the approximate master equation \cite{PhysRevA.45.8045}, one introduces the atomic transition and inversion operators between the dressed states,
\be
S_-=|-\ra\la+|,\quad S_+=|+\ra\la-|, \quad S_z=|+\ra\la+|-|-\ra\la-|.
\label{tran_dress}
\e
Using Eq. (\ref{tran_dress}), one can express the atomic operators of the bare states basis as
\beml
\begin{align}
\s_-&=S^-_F+S^-_R+S^-_T,\\
\s_+&=S^+_F+S^+_R+S^+_T,
\end{align}
\label{sS}
\eml
where the operators
\be
S^-_F=c^2_{\theta/2}S_-, \quad S^-_T=-s^2_{\theta/2}S_-, \quad S^-_R=s_{\theta/2}c_{\theta/2}S_z/2,
\e
describe the emission of photons into the fluorescence (F), Rayleigh (R), and three-photon (T) lines of the triplet (see Fig. \ref{dressed_levels}), with the account of the $\theta$-dependent weights of the corresponding processes. Using  the representation (\ref{sS}), and employing the {\it secular} approximation \cite{cohen_tannoudji_API} wherein interference between the emission processes down the dressed states giving rise to different peaks of the triplet is ignored, it is possible to reduce the dissipative part of Eq. (\ref{meq}) to an incoherent sum of spontaneous decay processes into the three components of the RF triplet \cite{PhysRevA.45.8045}. The temporal normalized second-order intensity correlation functions of photons transmitted by two wide spectral filters ($\gamma\ll\Gamma\ll \Omega$), tuned to the components $\alpha,\beta$ of the triplet ($\alpha,\beta=$F, R, T, which means that the filters' resonance frequencies coincide 
with the positions of the peaks of the RF triplet: $\delta_1, \delta_2=-\Omega, 0, \Omega$; see Fig. \ref{dressed_levels}), can then be found analytically \cite{PhysRevA.45.8045} (see Appendix \ref{app:analytics}). \begin{widetext}
\begin{center}
\begin{figure} 
{\includegraphics[width=14cm]{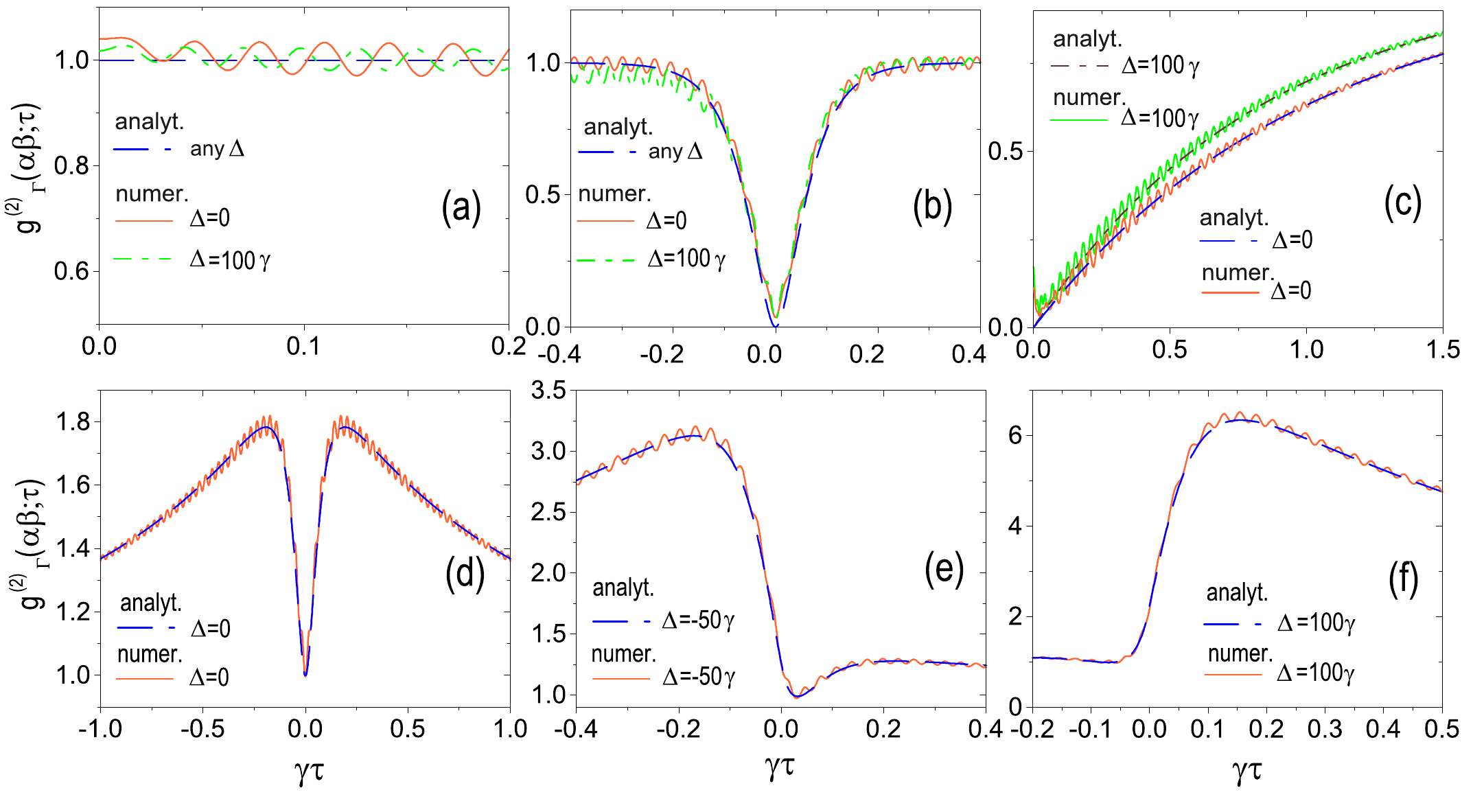}}
\caption{(Color online) 
 Temporal normalized second-order intensity correlation function, $g^{(2)}_\Gamma(\alpha\beta;\tau)$, for the components $\alpha$, $\beta=F, R, T$ of the RF triplet (see Fig. \ref{dressed_levels}) that are resolved by the interference filter(s) with $\Gamma=20\gamma$ in the limit of well separated spectral lines ($v=200\gamma$). Our numerical results (``numer.'') are plotted for exact resonance ($\Delta=0$) and for the detuned driving (see legends) along with the analytical solutions (``analyt.''), derived in \cite{PhysRevA.45.8045} (see Appendix \ref{app:analytics}). (a) $\alpha\beta=$RR, analyt.= Eq. (\ref{g2RR}); (b)  $\alpha\beta=$RF=RT, analyt.= Eq. (\ref{g2TR}); (c) $\alpha\beta=$FF=TT, analyt.= Eq. (\ref{g2TT}); (d), (e) $\alpha\beta=$TF, analyt.=Eqs. (\ref{g2TF}), (\ref{g2FT}); (f) $\alpha\beta=$FT, analyt.= Eqs. (\ref{g2TF}), (\ref{g2FT}). Note that analytical results are independent of $\Delta$ in plots (a) and (b). Further on, plots (a) and (c) are presented only for $\tau\geq 0$, since the function $g^{(2)}_\Gamma(\alpha\alpha;\tau)$ is time symmetric for $\alpha=$F, R, T.}
\label{fig:g2t}
\end{figure}
\end{center}
\end{widetext}
In Fig. \ref{fig:g2t}, along with the approximate results obtained in \cite{PhysRevA.45.8045}, we present our results for the normalized temporal second-order intensity correlation function, 
\be
g^{(2)}_\Gamma(\alpha\beta;\tau)\equiv\frac{G^{(2,2)}_\tau(\Gamma;\alpha,\beta)}{G^{(1,1)}_0(\Gamma;\alpha)G^{(1,1)}_0(\Gamma;\beta)},
\label{def_g2twell}
\e
where the numerator and denominator in the right hand side of Eq. (\ref{def_g2twell}) are given by Eqs. (\ref{def_G22tau}) and (\ref{def_G11}), respectively. 
\begin{figure} 
{\includegraphics[width=8cm]{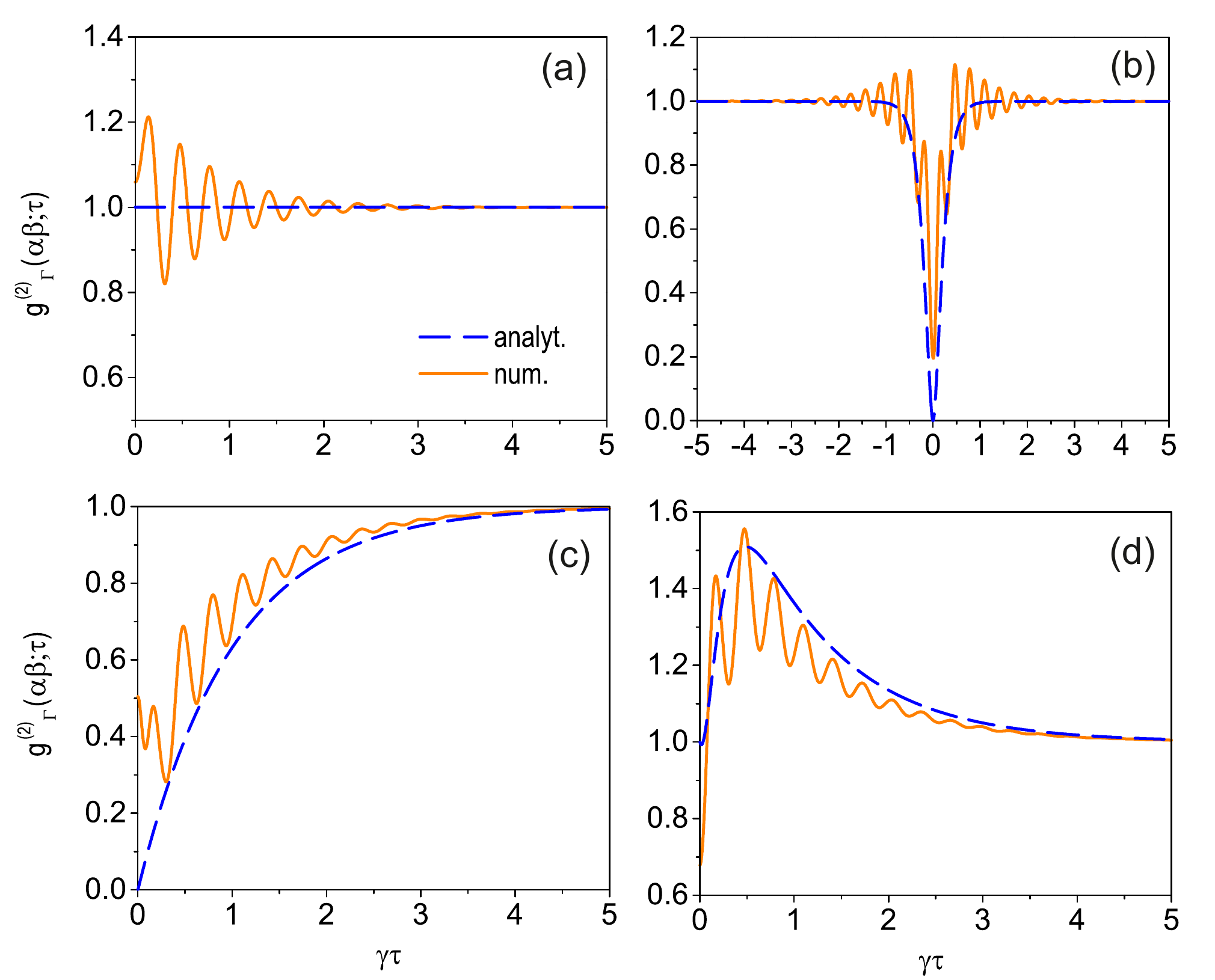}}
\caption{
(Color online) Same as in Fig.~\ref{fig:g2t} but for $v=20\gamma$, $\Delta=0$, and $\Gamma=6\gamma$. (a) $\alpha\beta=$RR, analyt.= Eq. (\ref{g2RR}); (b)  $\alpha\beta=$RF=RT, analyt.= Eq. (\ref{g2TR}); (c) $\alpha\beta=$FF=TT, analyt.= Eq. (\ref{g2TT}); (d) $\alpha\beta=$TF, analyt.= Eq. (\ref{g2TF}).}
\label{fig:g2t1}
\end{figure}

In all cases, the approximate analytical results of \cite{PhysRevA.45.8045} are very close to the exact behavior (see Fig. \ref{fig:g2t}). The only feature that is not captured within the approximate treatment are the oscillations of $g^{(2)}_\Gamma(\alpha\beta;\tau)$ with the frequency $\Omega$ and amplitude $\sim \gamma/\Omega\ll 1$. These oscillations arise due to interference between the emission processes giving rise to different lines of the RF spectrum. The interference effect is very small when the filters are tuned to the peaks of the triplet (the small amplitude of the oscillations is consistent with the error due to the secular approximation \cite{cohen_tannoudji_API,scully}). However, setting the filters's resonance frequencies in between the Rayleigh peak and either of the sidebands enhances the interference effect. In particular, interference between different spontaneous emission processes down the dressed state ladder results in the inversion of the reduced atomic state following detection of the frequency filtered photon \cite{Shatokhin01}. Another situation where interference between different emission processes cannot be ignored occurs beyond the limit of well-separated spectral lines. For instance, a decrease of the Rabi frequency down to $\Omega=20\gamma$, see Fig.~\ref{fig:g2t1}, results in the increase of the amplitude of the interferential oscillations of the function $g^{(2)}_\Gamma(\alpha\beta;\tau)$ and in the overall significant deviations of the exact behavior thereof from the predictions of \cite{PhysRevA.45.8045}.

Let us briefly remind the main properties of the function  $g^{(2)}_\Gamma(\alpha\beta;\tau)$ that are manifest in Fig. \ref{fig:g2t}. 
Namely, the photons within the Rayleigh line exhibit the Poisson statistics, that is, they are uncorrelated [$g^{(2)}_\Gamma(RR;0)\approx 1$, see Fig. \ref{fig:g2t}(a)]. In contrast, the photons emitted into the sidebands, as well as the photons from the central peak and either of the sidebands exhibit antibunching  [$g^{(2)}_\Gamma(TT;0)=g^{(2)}_\Gamma(FF;0)=g^{(2)}_\Gamma(RT;0)=g^{(2)}_\Gamma(RF;0)\approx 0$, see Fig. \ref{fig:g2t}(b,c)]. Finally, the photons from different sidebands are uncorrelated at $\Delta=0$ [see Fig. \ref{fig:g2t}(d)]. Nonzero detunings $\Delta$ lead to the asymmetry of the time-delayed coincidence rate and to bunching of photons from different sidebands [$g^{(2)}_\Gamma(FT;0)>1$, see Fig. \ref{fig:g2t}(e,f)]. The asymmetry of the function $g^{(2)}_\Gamma(FT;\tau)$ is a manifestation of a definite time order between the processes giving rise to the detected photons -- fluorescence (F) occurs after the three-photon (T) scattering process, in agreement with \cite{PhysRevLett.45.617}.

\subsubsection{The function $g^{(2)}_\Gamma(\delta_1,\delta_2;0)$ and ``leapfrog'' transitions}
\label{leapfrog}
The results of Sec. \ref{secular} suggest that the zero-delay coincidence rate $g^{(2)}_\Gamma(\alpha\beta;0)$ allows one to distinguish between three different kinds of statistics: Poisson, bunching, and antibunching. The type of the statistics stems from the dependence of photon correlations on the particular two-photon emission cascade down the dressed state ladder \cite{apanasevich79,Cohen-Tannoudji79}. Therefore, in a certain sense, the function $g^{(2)}_\Gamma(\alpha\beta;0)$ reflects information about the elementary scattering processes on a laser driven atom.

This fact has encouraged some authors to consider the function $g^{(2)}_\Gamma(\delta_1,\delta_2;0)$, where $\delta_1$ and $\delta_2$ are arbitrary, as a quantity identifying possible scattering processes \cite{PhysRevLett.109.183601,tudela13}. Examples of the normalized correlation function $g^{(2)}_\Gamma(\delta_1,\delta_2;0)$ are presented in Fig.~\ref{fig:g2_0}. This function exhibits the mirror reflection symmetry about the diagonal $\delta_1=\delta_2$ and -- at $\Delta=0$ -- about the antidiagonal $\delta_1=-\delta_2$, in agreement with Eqs. (\ref{sym_diag},\ref{sym_antidiag}). 
However, the most prominent feature of $g^{(2)}_\Gamma(\delta_1,\delta_2;0)$ is its very large values $\gg 1$ (`resonances') for the values of $\delta_1$, $\delta_2$ that lie outside the positions of the maxima of the RF spectrum. The origin of these resonances has been attributed to a special class of elementary scattering processes termed {\it leapfrog} transitions \cite{tudela13}. 
According to \cite{tudela13}, these transitions cannot be described as emission cascades down the dressed states' ladder (see Fig.~\ref{dressed_levels}), but occur via two-photon jumps mediated by virtual states.  Recently, strong correlations of the function $g^{(2)}_\Gamma(\delta_1,\delta_2;0)$ in the domains of $\delta_1, \delta_2$-values predicted in \cite{tudela13} have been measured in  \cite{peiris15} and regarded as the experimental evidence of the leapfrog transitions.  

\begin{widetext}
\begin{center}
\begin{figure} 
{\includegraphics[width=13cm]{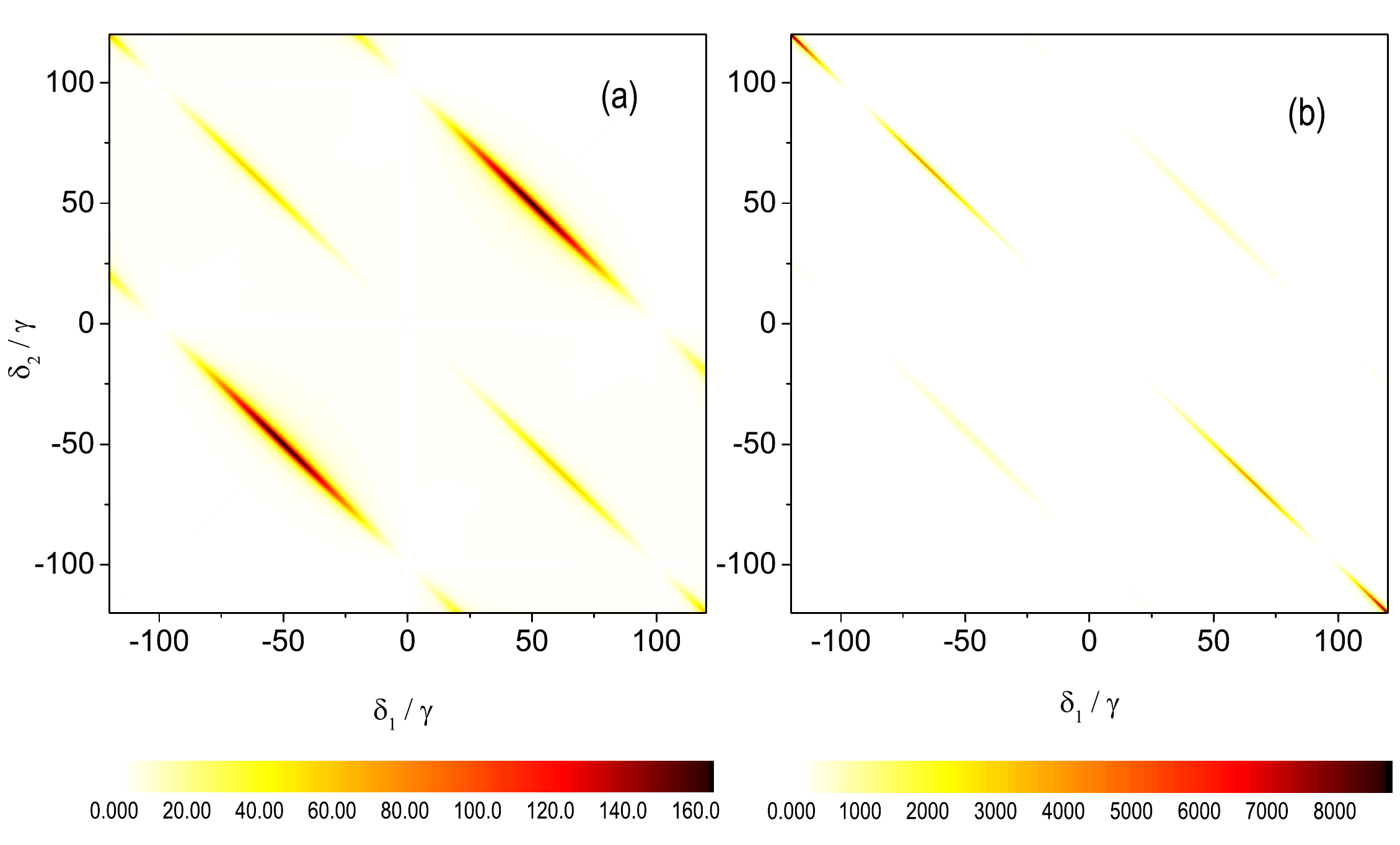}}
\caption{
(Color online) Contour plots of the normalized correlation function $g^{(2)}_\Gamma(\delta_1,\delta_2;0)$ in the limit of well separated spectral lines, at $\Gamma=0.4\gamma$, and (a) $v=100\gamma$, $\Delta=0$; (b) $v=50\gamma$, $\Delta=80\gamma$. Both plots exhibit strong correlations at the values of $\delta_1$, $\delta_2$ that lie {\it beyond} the positions the peaks of the Mollow triplet at (a) $\delta_1,\delta_2=0,\pm 100\gamma$, (b) $\delta_1,\delta_2=0,\pm 94.34\gamma$.}
\label{fig:g2_0}
\end{figure}
\end{center}
\end{widetext}
However, the above interpretation of the function $g^{(2)}_\Gamma(\delta_1,\delta_2;0)$ outstretches the physical meaning of this quantity. Indeed, it is the emission spectrum that shows which of the scattering processes are possible, whereas the spectral intensity correlation function identifies which of them are correlated \cite{cohen_tannoudji_API}. Hence, the maxima of the function $g^{(2)}_\Gamma(\delta_1,\delta_2;0)$ should not be associated with a new kind of scattering processes. 

Therefore, we would like to present an alternative explanation of the behavior of the function $g^{(2)}_\Gamma(\delta_1,\delta_2;0)$ in Fig. \ref{fig:g2_0}. The resonances in Fig.~\ref{fig:g2_0} can be understood using the dressed state picture, if we recall that $g^{(2)}_\Gamma(\delta_1,\delta_2;0)$ is the {\it normalized} correlation function. Its very large values are attained on the tails of the Lorentzian distribution (i.e. for $|\delta_{1,2}-\omega_M|\gtrsim 10\gamma$, with $\omega_M=0,\pm \Omega$),  where the denominator of Eq.~(\ref{def_g2twell}) (the product of spectral intensities) is very small. Yet the intensity correlation function [the numerator of Eq.~(\ref{fig:g2_0})] can be {\it relatively} large, resulting in the magnitude of the ratio $\gg 1$. Such a condition is realized, for example, for pairs of the transmitted photons whose frequencies satisfy the energy conservation relation, $\omega_1+\omega_2=2\omega_L$ [i.e., $\delta_1+\delta_2=0$, which corresponds to the main antidiagonals $\delta_1=-\delta_2$ in Fig. \ref{fig:g2_0}(a,b)]. Other domains of strong correlations in Fig. \ref{fig:g2_0} lie along the lateral antidiagonals $\delta_1=-\delta_2\pm \Omega$. In this case the normalized correlation function attains the maximum values on the crossings with the main diagonal; that is, at $\delta_1=\delta_2=\pm \Omega/2$, which lie in between the central peak and one of the sidebands. In this case, interference between the processes giving rise to the Rayleigh peak and to either of the sideband resonances of the Mollow triplet, comes into play. 
Thus, such properties as two-photon entanglement, violations of classical inequalities, etc., are not due to special virtual transitions \cite{PhysRevA.90.052111,PhysRevA.91.043807,PhysRevLett.115.196402} but rather due to {\it frequency (post)selection} on the tails of the spectral distribution of the light resonantly scattered by an atom.

Having thus shed light on the behavior of the function $g^{(2)}_\Gamma(\delta_1,\delta_2;0)$, we will next present a true measure of spectral correlations in RF which exhibits pronounced resonances only in the frequency domains that coincide with positions of the peaks of the RF spectrum. 

\subsection{A true measure of spectral correlations in resonance fluorescence}
\label{sec:unnormalized}
To characterize spectral correlations in RF, instead of $g^{(2)}_\Gamma(\delta_1,\delta_2;0)$ we will use the {\it unnormalized} function \cite{Apanasevich197783}
\be
\Delta G^{(2)}\!(\Gamma;\delta_1,\delta_2)\!\equiv \!G^{(2,2)}_0\!(\Gamma;\delta_1,\delta_2)-G^{(1,1)}_0\!(\Gamma;\delta_1)G^{(1,1)}_0\!(\Gamma;\delta_2).
\label{def_dG}
\e
By definition, the function $\Delta G^{(2)}(\Gamma;\delta_1,\delta_2)$ possesses the symmetry properties (\ref{sym_diag}) and (\ref{sym_antidiag}). Furthermore, 
this function has the following meaning: it attains positive (negative) values for correlated (anticorrelated) pairs of spectrally filtered photons and it vanishes for uncorrelated pairs thereof. It was predicted in \cite{Apanasevich197783} that in the limit $\Gamma\to 0$ the sideband photons satisfying the condition $\omega_1+\omega_2=2\omega_L$ are strongly correlated, but the function (\ref{def_dG}) has not been systematically studied. 

In this work we illustrate Eq. (\ref{def_dG}) in Fig. \ref{fig:Om=100_D=0} for the same parameters' values as used in Fig. \ref{fig:g2_0}. It is clear that the location of the resonances in Figs. \ref{fig:g2_0} and \ref{fig:Om=100_D=0} are complementary to each other: In Fig. \ref{fig:Om=100_D=0} their position in the $(\delta_1,\delta_2)$ plane coincides with the position of the peaks of the RF triplet. Outside these regions of pronounced (anti)correlations (which for $\Gamma\lesssim\gamma$ spread over areas with a linear size $\sim\gamma$), the spectral correlation function  in Fig.~\ref{fig:Om=100_D=0} forms a background where the absolute value of $\Delta G^{(2)}(\Gamma;\delta_1,\delta_2)$ is several orders of magnitude smaller than that at the peaks. 

\begin{widetext}
\begin{center}
\begin{figure} 
{\includegraphics[width=13cm]{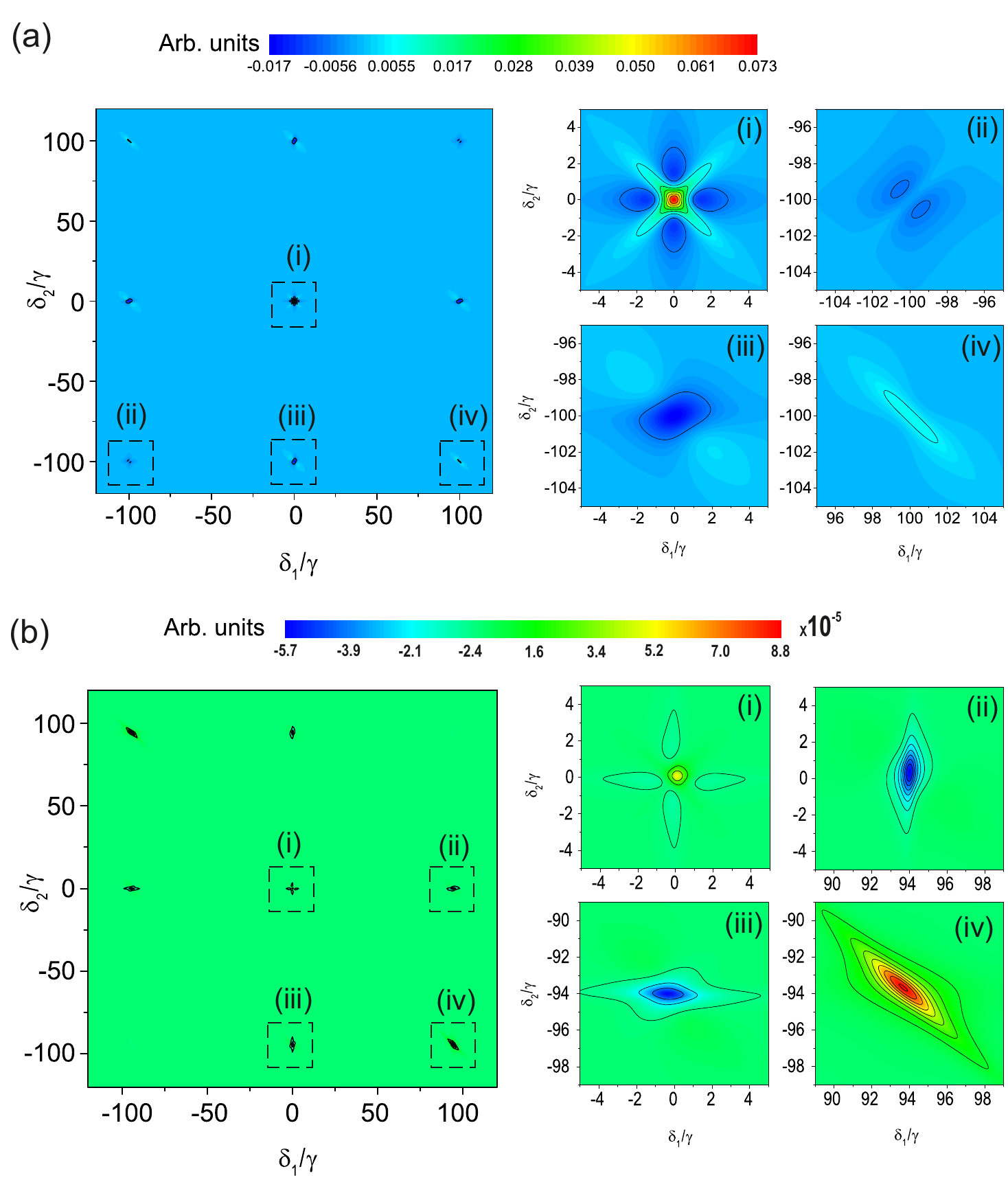}}
\caption{
(Color online) Contour plots of the spectral correlation function $\Delta G^{(2)}(\Gamma;\delta_1,\delta_2)$ at $\Gamma=0.4\gamma$ and (a) $v=100\gamma$, $\Delta=0$, and (b) $v=50\gamma$, $\Delta=80\gamma$. Left panels show that $\Delta G^{(2)}(\Gamma;\delta_1,\delta_2)$ exhibits (a) nine and (b) seven resonances. Four of the resonances in (a) and (b) are framed in dashed boxes and labeled by roman numerals i,ii,iii, and iv. Right panels show magnified resonances (i)-(iv). The structure of the remaining resonances can be extracted using the symmetry properties: (a), (b) $\Delta G^{(2)}(\Gamma;\delta_1,\delta_2)=\Delta G^{(2)}(\Gamma;\delta_2,\delta_1)$ [see Eq. (\ref{sym_diag})] and (a) $\Delta G^{(2)}(\Gamma;\delta_1,\delta_2)=\Delta G^{(2)}(\Gamma;-\delta_1,-\delta_2)$ [see Eq. (\ref{sym_antidiag}))].}
\label{fig:Om=100_D=0}
\end{figure}
\end{center}
\end{widetext}

Finally, let us discuss the character of spectral correlations featured in Fig. \ref{fig:Om=100_D=0}. We remind that a variation of the filters' bandwidths can lead to a modification of the photon statistics \cite{Armstrong:66,1464-4266-2-2-317}. In other words, the statistics of the fields that are valid for relatively narrow filters, with $\Gamma\leq \gamma$, in general differ  from the statistics in the case $\Gamma\gg \gamma$, discussed in Sec.~\ref{secular} (note also Fig.~\ref{fig:g2t} at $\tau=0$). Figures \ref{fig:Om=100_D=0}(a) panels (i,iv) and \ref{fig:Om=100_D=0}(b) panels (i,iv)]  feature strong correlations for the filters tuned either both to the Rayleigh peak or to the opposite sidebands. Furthermore, within the Rayleigh lines, filters tuned symmetrically with respect to the laser frequency and filters tuned to the same frequency are also correlated. The former correlation results from the energy conservation for photons satisfying the relation $\delta_1+\delta_2=0$; the latter one has a purely classical origin, since $\Delta G^{(2)}(\Gamma;\delta,\delta)$ has the meaning of the variance. A signature of the latter (classical) correlation can be noticed also when both filters are tuned to the sidebands [see Fig.~\ref{fig:Om=100_D=0}(a), panel (ii)]. The remaining (negative) resonances emerge when one of the filters is tuned to the central peak and another one to either of the sidebands [see Fig.~\ref{fig:Om=100_D=0}(a), panel (iii); (b), panels (ii,iii)]. Also, photons from the central peak are anticorrelated when the filters are tuned asymmetrically with respect the laser frequency [see Figs.~\ref{fig:Om=100_D=0}(a) and \ref{fig:Om=100_D=0}(b), panel (i)]. The given summary is consistent with the results of \cite{0295-5075-21-3-006}, where temporal correlations between the fields, passed through narrow filters that were tuned within the components of the RF triplet, have been analyzed.  

\section{Conclusion}
\label{sec:disc}
We developed an efficient method to calculate spectral correlation functions in single-atom resonance fluorescence. Our method represents a generalization of the so-called signal processing approach \cite{nienhuis84,0022-3700-20-18-027,PhysRevA.42.503} to spectral filtration -- wherein an interferometer is treated as a black box relating the output to the input by the response function -- to arbitrary parameters of the filters and laser driving field. An appealing feature of our method is an intuitive character of its diagrammatic implementation and the possibility to derive general expressions for the correlation functions in an analytical form. 
 
In this work we applied our method to assess spectral correlations in the limit of well separated spectral lines of the RF spectrum and restricted ourselves to the 
second-order intensity correlation signals passed through the Fabry-Perot interferometers with equal bandwidths. Thereby we, on the one hand, checked the validity of our results by comparing them with the ones obtained previously \cite{PhysRevA.45.8045} in the secular limit. On the other hand, we showed that interference effects between the contributions to the different components of the RF triplet are not entirely negligible.   

Finally, we critically examined the concept of ``leapfrog'' transitions \cite{tudela13}. We showed that large values of the normalized spectral intensity correlation function and associated effects reported in \cite{tudela13,PhysRevA.90.052111} can be understood as a result of spectral post-selection. Moreover, we explored a true measure of spectral correlations and showed that its behavior can fully be understood by considering the spontaneous two-photon cascades down the dressed states ladder, without introducing any new kind of transitions.

It would be interesting to apply our present method to study higher-order spectral correlations of RF in future work. We would also like to clarify the connection between the spectral correlation functions in RF with spectral resolution and the single atom spectral response functions that appear in the theory of multiple scattering of light by atoms \cite{shatokhin12b}.

\acknowledgements
V.N.S. thanks A. Buchleitner for stimulating discussions, and he acknowledges support through the EU Collaborative project QuProCS (Grant Agreement No. 641277). S.Ya.K. acknowledges support of Belarusian Republican Foundation for Fundamental Research (Grant No. F16KOR2-001) and of the State Scientific Programme ``Convergence 2020.''

\appendix
\section{Green functions}
\label{green_a}
Here we provide the explicit expressions for the Laplace transforms of the  atomic Green's matrix elements which are needed to determine matrices $\tilde{\bf D}^{[+]}(p)$ and $\tilde{\bf D}^{[-]}(p)$ [see Eq. (\ref{Dpm})]
\beml
\begin{align}
\tilde{\cal D}^{11}_{12}(p)&=[\tilde{\cal D}^{11}_{21}(p)]^*\n\\&=-\frac{i(p+2\gamma)(p+\gamma-i\Delta)v}{2pQ(p)},\\
\tilde{\cal D}^{11}_{22}(p)&=\frac{(p+\gamma)v^2}{2pQ(p)},\\
\tilde{\cal D}^{21}_{12}(p)&=\tilde{\cal D}^{12}_{21}(p)=\frac{v^2}{2Q(p)},\\
\tilde{\cal D}^{21}_{21}(p)&=[\tilde{\cal D}^{12}_{12}(p)]^*\n\\&=\frac{2(p+2\gamma)(p+\gamma+i\Delta)+v^2}{2Q(p)},\\
\tilde{\cal D}^{21}_{22}(p)&=[\tilde{\cal D}^{12}_{22}(p)]^*=-\frac{i(p+\gamma+i\Delta)v}{2Q(p)},
\end{align}
\label{eltsD}
\eml
where 
$Q(p)=(p+2\gamma)[\Delta^2+(p+\gamma)^2]+(p+\gamma)v^2$.
 It can be shown \cite{PhysRev.188.1969} that for an arbitrary detuning $\Delta$, the three roots of the polynomial $Q(p)$ have negative real parts; these roots are either all real or one of them is real, whereas the remaining two are complex conjugates of each other. At exact resonance ($\Delta=0$) the roots of $Q(p)$ are
$p_0=-\gamma$, $p_{\pm}=-3\gamma/2\pm i\Omega$,
where $\Omega=\sqrt{v^2-\gamma^2/4}$ is the modified Rabi frequency. 

Let us finally present also the steady-state solution ${\bf r}_\infty=\lim_{t\to\infty}{\bf r}(t)$ for the vector ${\bf r}(t)=[\rho_{12}(t),\rho_{21}(t),\rho_{22}(t)]^T$:
\beml
\begin{align}
\rho_{12}(\infty)&=\frac{-i(\gamma-i\Delta) v}{2\lt(\gamma^2+\Delta^2\rt)+ v^2},\\
\rho_{21}(\infty)&=[\rho_{12}(\infty)]^*,\\
\rho_{22}(\infty)&=\frac{ v^2}{4\lt(\gamma^2+\Delta^2\rt)+v^2}.
\end{align}
\label{rho_inf}
\eml

\section{Derivation of Eq. (\ref{aux0})}
\label{conv1}
To take the convolution integrals in Eq. (\ref{int_t}), we introduce new integration variables, $x_1, x_2, \ldots, x_{n+m}$, which are related to the old ones through
\be
t_{j_k}=t-\sum_{l=1}^k x_l \quad (k=1,\ldots, n+m). 
\label{sub1}
\e
It is easy to check that the Jacobian of this transformation, $J=|\partial(t_{j_1},\ldots, t_{j_{n+m}})/\partial(x_1,\ldots, x_{n+m})|=1$. In new variables, the right-hand side of Eq. (\ref{int_t}) transforms to
\begin{widetext}
\begin{align}
G^{(n,m)}&=
\sum_{\pi(j_1,\ldots,j_{n+m})}\lim_{t\to\infty}
\int_0^tdx_1\int_0^{t-x_1}dx_2
\ldots \int_0^{t-x_1-\ldots-x_{n+m-1}}dx_{n+m}\n\\
&\times \prod_{k=1}^{n+m}\Gamma_{j_k} e^{-\lambda_{j_k}\sum_{l=1}^kx_l}
\lt\{{\bf D}^{[s_{j_2}]}(x_2)\ldots {\bf D}^{[s_{j_{n+m}}]}(x_{n+m}){\bf r}(t-\sum_{l=1}^{n+m}x_l)\rt\}_{s_{j_1}}
\n\\
&=\sum_{\pi(j_1,\ldots,j_{n+m})}
\int_0^\infty dx_1e^{-\Lambda_1x_1}\int_0^{\infty}dx_2e^{-\Lambda_2x_2}
\ldots \int_0^{\infty}dx_{n+m}e^{-\Lambda_{n+m}x_{n+m}}\n\\
&\times \prod_{k=1}^{n+m}\Gamma_{j_k}\lt\{{\bf D}^{[s_{j_2}]}(x_2)\!\ldots\! {\bf D}^{[s_{j_{n+m}}]}(x_{n+m}){\bf r}_\infty\rt\}_{s_{j_1}}
\!\!=\!\frac{1}{\Lambda_1}\sum_{\pi(j_1,\ldots,j_{n+m})}\!\!\Gamma_{j_1}\!\lt\{\lt[\prod_{k=2}^{n+m}\Gamma_{j_k}\tilde{\bf D}^{[s_{j_k}]}(\Lambda_k)\rt]\!{\bf r}_\infty\!\rt\}_{s_{j_1}}\!,
\label{aux2}
\end{align}
where  ${\bf r}_\infty$ is given by Eq. (\ref{rho_inf}), 
\be
\Lambda_k=\sum_{l=k}^{n+m}\lambda_{j_l},
\e
with $\lambda_{j_l}=\Gamma_{j_l}+i\delta_{j_l}$, and $\tilde{\bf D}^{[\pm]}(p)$ is Laplace transform of the propagator ${\bf D}^{[\pm]}(t)$:
\be
\tilde{\bf D}^{[\pm]}(p)=\int_0^\infty dt e^{-pt}{\bf D}^{[\pm]}(t), \quad \re[p]\geq 0.
\e
The elements of the matrices $\tilde{\bf D}^{[\pm]}(p)$ are given in Eq. (\ref{eltsD}).

\section{Temporal correlation functions of spectrally resolved photons}
\label{time_delay}
We begin with the calculation of $I_2(\tau)$. Expanding the right-hand side of Eq. (\ref{i2}) into fully time-ordered integrals, we obtain
\begin{align}
I_{2}(\tau)&=\lim_{t\to\infty}\sum_{\pi(j_1,j_2,j_3)}\int_t^{t+\tau}dt_2\int_0^tdt_{j_1}\int_0^{t_{j_1}}dt_{j_2}\int_0^{t_{j_2}}dt_{j_3}
\Gamma^4 e^{-2\Gamma\tau}\prod_{k=1}^4e^{-\lambda_k(t-t_k)}\n\\
&\times \{{\bf D}^{[s_{j_1}]}(t_2-t_{j_1}){\bf D}^{[s_{j_2}]}(t_{j_1}-t_{j_2})
{\bf D}^{[s_{j_3}]}(t_{j_2}-t_{j_3}){\bf r}(t_{j_3})\}_+,
\end{align}
where $j_1,j_2,j_3\in\{1,3,4\}$. 
After the transformation of variables
\be
t_2=t+\tau-x_1, \quad t_{j_k} = t+\tau-\sum_{l=1}^{k+1} x_l
\label{t_shift}
\e
that, up to a time shift $\tau$, coincides with (\ref{sub1}) (hence, its Jacobian $|J|=1$),
 we arrive at
\begin{align}
I_{2}(\tau)&=
\lim_{t\to\infty}\sum_{\pi(j_1,j_2,j_3)}\int_0^{\tau}dx_1\int_{\tau-x_1}^{t+\tau-x_1}dx_2\int_0^{t+\tau-x_1-x_2}dx_3\int_0^{t+\tau-x_1-x_2-x_3}dx_4\n\\
&\times \Gamma^4 e^{2\Gamma\tau}e^{-\lambda_2x_1}\prod_{k=1}^3e^{-\lambda_{j_k}\sum_{l=1}^{k+1}x_l}
\lt\{{\bf D}^{[s_{j_1}]}(x_2){\bf D}^{[s_{j_2}]}(x_3)
{\bf D}^{[s_{j_3}]}(x_4){\bf r}(t+\tau-\sum_{l=1}^4x_l)\rt\}_+\n\\
&=\Gamma^4 e^{2\Gamma\tau}\sum_{\pi(j_1,j_2,j_3)}\int_0^{\tau}dx_1\int_{\tau-x_1}^{\infty}dx_2\int_0^{\infty}dx_3\int_0^{\infty}dx_4\n\\
&\times e^{-(\lambda_2+\sum_{k=1}^3\lambda_{j_k})x_1}e^{-\sum_{k=1}^3\lambda_{j_k}x_2}e^{-(\lambda_{j_2}+\lambda_{j_3})x_3}
e^{-\lambda_{j_3}x_4}
\lt\{{\bf D}^{[s_{j_1}]}(x_2){\bf D}^{[s_{j_2}]}(x_3)
{\bf D}^{[s_{j_3}]}(x_4){\bf r}_\infty\rt\}_+\n\\
&\!=\!\Gamma^4 e^{2\Gamma\tau}\!\!\sum_{\pi(j_1,j_2,j_3)}\int_0^{\tau}\!dx_1e^{-4\Gamma x_1}\int_{\tau-x_1}^{\infty}\!dx_2e^{-\sum_{k=1}^3\lambda_{j_k}x_2}
\lt\{{\bf D}^{[s_{j_1}]}(x_2)\tilde{\bf D}^{[s_{j_2}]}(\lambda_{j_2}\!+\!\lambda_{j_3})
\tilde{\bf D}^{[s_{j_3}]}(\lambda_{j_3}){\bf r}_\infty\rt\}_+,
\label{final_I2}
\end{align}
where we have used the identity $\lambda_2+\sum_{k=1}^3\lambda_{j_k}=4\Gamma$. Thus, the calculation of $I_2(\tau)$ involves a double integration and requires the expression for the time-dependent propagators ${\bf D}^{[\pm]}(t)$. The latter can easily be found by the inverse Laplace transform of  $\tilde{\bf D}^{[\pm]}(p)$. Namely, each element of ${\bf D}^{[\pm]}(t)$ represents a sum of decaying exponentials  (see Appendix \ref{green_a}), such that taking the integrals in (\ref{final_I2}) is elementary. 

In full analogy with the above result, for $I_3(\tau)$ we obtain,
\begin{equation}
I_{3}(\tau)\!=\!\Gamma^4 e^{2\Gamma\tau}\sum_{\pi(i_1,i_2,i_3)}\int_0^{\tau}dx_1e^{-4\Gamma x_1}\int_{\tau-x_1}^{\infty}dx_2e^{-\sum_{k=1}^3\lambda_{i_k}x_2}
\lt\{{\bf D}^{[s_{i_1}]}(x_2)\tilde{\bf D}^{[s_{i_2}]}(\lambda_{i_2}+\lambda_{i_3})
\tilde{\bf D}^{[s_{i_3}]}(\lambda_{i_3}){\bf r}_\infty\rt\}_-,
\end{equation}
where $i_1, i_2, i_3\in{1,2,4}$.

Finally, we calculate $I_4(\tau)$. By definition,
\begin{align}
I_{4}(\tau)&=\lim_{t\to\infty}\sum_{\pi(i_1,i_2),\pi(j_1,j_2)}\int_t^{t+\tau}dt_{i_1}\int_t^{t_{i_1}}dt_{i_2}\int_0^{t}dt_{j_1}\int_0^{t_{j_1}}dt_{j_2}\n\\
&\times \Gamma^4 e^{-2\Gamma\tau}\prod_{k=1}^4e^{-\lambda_k(t-t_k)}
\{{\bf D}^{[s_{i_2}]}(t_{i_1}-t_{i_2}){\bf D}^{[s_{j_1}]}(t_{i_2}-t_{j_1})
{\bf D}^{[s_{j_2}]}(t_{j_1}-t_{j_2}){\bf r}(t_{j_2})\}_{s_{i_1}},
\end{align}
where $i_1,i_2\in\{2,3\}$ and $j_1,j_2\in \{1,4\}$. Performing the transformation of variables 
\be
t_{i_k}=t+\tau-\sum_{l=1}^kx_l, \quad t_{j_k} = t+\tau-\sum_{l=1}^{k+2} x_l,
\label{t_shift2}
\e
which is similar to (\ref{t_shift}), and taking the limit $t\to\infty$, we obtain the result
\begin{align}
I_{4}(\tau)&=\Gamma^4 e^{2\Gamma\tau}\sum_{\pi(i_1,i_2),\pi(j_1,j_2)}\int_0^{\tau}dx_1e^{-4\Gamma x_1}\int_0^{\tau-x_1}dx_2e^{-(\lambda_{i_2}+2\Gamma)x_2}\int_{\tau-x_1-x_2}^{\infty}dx_3e^{-2\Gamma x_3}\n\\
&\times \{{\bf D}^{[s_{i_2}]}(x_2){\bf D}^{[s_{j_1}]}(x_3)
\tilde{\bf D}^{[s_{j_2}]}(\lambda_{j_2}){\bf r}_\infty\}_{s_{i_1}}.
\end{align}

\section{Analytical formulas for the function $g^{(2)}_\Gamma(\alpha\beta,\tau)$}
\label{app:analytics}
For reference, here we reproduce the analytical expressions for the functions $g^{(2)}_\Gamma(\alpha\beta;\tau)$ ($\alpha,\beta$=F, R, T) that  were derived on the basis of an approximate master equation in \cite{PhysRevA.45.8045}:
\beml
\begin{align}
g^{(2)}_\Gamma(RR;\tau)&\!=\!1,\label{g2RR}\\
g^{(2)}_\Gamma(FR;\tau)&\!=\!g^{(2)}_\Gamma(RF;\tau)=g^{(2)}_\Gamma(TR;\tau)=g^{(2)}_\Gamma(RT;\tau)\!=\!(1-e^{-\Gamma\tau})^2,\label{g2TR}\\
g^{(2)}_\Gamma(TT;\tau)&\!=\!g^{(2)}_\Gamma(FF;\tau)=1-e^{-\gamma_1\tau},\label{g2TT}\\
g^{(2)}_\Gamma(TF;\tau)&\!=\!\frac{c_{\theta/2}^4}{s_{\theta/2}^4}\lt(e^{-\gamma_1\tau}-1\rt)+\lt(1+\frac{c_{\theta/2}^4}{s^4_{\theta/2}}\rt)\lt(1-\frac{1}{2}e^{-\Gamma \tau}\rt)^2+\lt(1+\frac{s_{\theta/2}^4}{c^4_{\theta/2}}\rt)\frac{1}{4}e^{-2\Gamma \tau},\label{g2TF}\\
g^{(2)}_\Gamma(FT;\tau)&\!=\!\frac{s_{\theta/2}^4}{c_{\theta/2}^4}\lt(e^{-\gamma_1\tau}\!-\!1\rt)\!+\!\lt(\!1\!+\!\frac{s_{\theta/2}^4}{c^4_{\theta/2}}\rt)\lt(1\!-\!\frac{1}{2}e^{-\Gamma \tau}\rt)^2\!+\!\lt(1\!+\!\frac{c_{\theta/2}^4}{s^4_{\theta/2}}\rt)\frac{1}{4}e^{-2\Gamma \tau},\label{g2FT}
\end{align}
\eml
where $\gamma_1=2\gamma(c^4_{\theta/2}+s^4_{\theta/2})$ and $c_{\theta/2}$, $s_{\theta/2}$
are defined after Eq. (\ref{def_dressed}).
\end{widetext}
\bibliographystyle{prstil}
\bibliography{lit_spectral}

\begin{thebibliography}{10}

\bibitem{heitler}
W. Heitler, {\em The quantum theory of radiation}, 3rd ed. (Oxford University
  Press, ADDRESS, 1954).

\bibitem{cohen_tannoudji_API}
C. Cohen-Tannoudji, J. Dupont-Roc, and G. Grynberg, {\em Atom-Photon
  Interactions} (Wiley, New York, 1992).

\bibitem{apanasevich64}
P.~A. Apanasevich, Opt. Spectrosc. {\bf 16},  387  (1964).

\bibitem{PhysRev.188.1969}
B.~R. Mollow, Phys. Rev. {\bf 188},  1969  (1969).

\bibitem{schuda74}
F. Schuda, C.~R.~S. Jr., and M. Hercher, J. Phys. B {\bf 7},  L198  (1974).

\bibitem{walther76}
W. Hartig, W. Rasmussen, R. Schieder, and H. Walther, Z. Phys. A {\bf 278},
  205  .

\bibitem{grove77}
R.~E. Grove, F.~Y. Wu, and S. Ezekiel, Phys. Rev. A {\bf 15},  227  (1977).

\bibitem{Apanasevich197783}
P.~A. Apanasevich and S.~J. Kilin, Phys. Lett. A {\bf 62},  83  (1977).

\bibitem{Eberly:77}
J.~H. Eberly and K. W\'{o}dkiewicz, J. Opt. Soc. Am. {\bf 67},  1252  (1977).

\bibitem{Born_Wolf}
M. Born and E. Wolf, {\em Principles of Optics}, 7th ed. (Cambridge University
  Press, ADDRESS, 1999).

\bibitem{Shatokhin2000157}
V. Shatokhin and S. Kilin, Opt. Commun. {\bf 174},  157  (2000).

\bibitem{cohen-tannoudji77}
C. Cohen-Tannoudji and S. Reynaud, J. Phys. B {\bf 10},  345  (1977).

\bibitem{apanasevich79}
P.~A. Apanasevich and S.~J. Kilin, J. Phys. B {\bf 12},  L83  (1979).

\bibitem{dalibard83}
J. Dalibard and S. Reynaud, J. Physique {\bf 44},  1337  (1983).

\bibitem{Cresser198347}
J.~D. Cresser, Phys. Rep. {\bf 94},  47  (1983).

\bibitem{nienhuis84}
H.~F. Arnoldus and G. Nienhuis, J. Phys. B {\bf 17},  963  (1984).

\bibitem{0022-3700-20-18-027}
J.~D. Cresser, J. Phys. B {\bf 20},  4915  (1987).

\bibitem{PhysRevA.42.503}
L. Kn\"oll, W. Vogel, and D.-G. Welsch, Phys. Rev. A {\bf 42},  503  (1990).

\bibitem{PhysRevLett.67.2443}
C.~A. Schrama {\it et~al.}, Phys. Rev. Lett. {\bf 67},  2443  (1991).

\bibitem{PhysRevA.45.8045}
C.~A. Schrama {\it et~al.}, Phys. Rev. A {\bf 45},  8045  (1992).

\bibitem{0295-5075-21-3-006}
G. Nienhuis, Europhys. Lett. {\bf 21},  285  (1993).

\bibitem{PhysRevLett.45.617}
A. Aspect {\it et~al.}, Phys. Rev. Lett. {\bf 45},  617  (1980).

\bibitem{Nick-Vamivakas:2009fk}
A. Nick~Vamivakas, Y. Zhao, C.-Y. Lu, and M. Atature, Nat. Phys. {\bf 5},  198
  (2009).

\bibitem{Flagg:2009}
E.~B. Flagg {\it et~al.}, Nat. Phys. {\bf 5},  203  (2009).

\bibitem{Astafiev12022010}
O. Astafiev {\it et~al.}, Science {\bf 327},  840  (2010).

\bibitem{Pigeau:2015rw}
B. Pigeau {\it et~al.}, Nat. Commun. {\bf 6},  8603  (2015).

\bibitem{toyli16}
D.~M. Toyli {\it et~al.}, Phys. Rev. X {\bf 6},  031004  (2016).

\bibitem{UlhaqA.:2012ij}
A. Ulhaq {\it et~al.}, Nat. Photon. {\bf 6},  238  (2012).

\bibitem{PhysRevLett.109.183601}
E. del Valle {\it et~al.}, Phys. Rev. Lett. {\bf 109},  183601  (2012).

\bibitem{tudela13}
A. Gonz\'alez-Tudela {\it et~al.}, New J. Phys. {\bf 15},  033036  (2013).

\bibitem{PhysRevA.90.052111}
C. S\'anchez Mu\~noz, E. del Valle, C. Tejedor, and F.~P. Laussy, Phys. Rev. A
  {\bf 90},  052111  (2014).

\bibitem{PhysRevA.91.043807}
A. Gonz\'alez-Tudela, E. del Valle, and F.~P. Laussy, Phys. Rev. A {\bf 91},
  043807  (2015).

\bibitem{peiris15}
M. Peiris {\it et~al.}, Phys. Rev. B {\bf 91},  195125  (2015).

\bibitem{gardiner93}
C.~W. Gardiner, Phys. Rev. Lett. {\bf 70},  2269  (1993).

\bibitem{carmichael93}
H.~J. Carmichael, Phys. Rev. Lett. {\bf 70},  2273  (1993).

\bibitem{carmichael}
H.~J. Carmichael, {\em Statistical methods in quantum optics I: {M}aster
  equations and {F}okker-{P}lanck equations} (Springer Berlin / Heidelberg,
  ADDRESS, 2002).

\bibitem{apan78}
P. Apanasevich and S. Kilin, J. Appl. Spectr. {\bf 29},  931  (1978).

\bibitem{scully}
M.~O. Scully and M.~S. Zubairy, {\em Quantum optics} (Cambridge University
  Press, Cambridge, 1997).

\bibitem{glauber}
R.~J. Glauber, {\em Quantum theory of optical coherence} (Wiley-VCH, ADDRESS,
  2007).

\bibitem{Knoll:86}
L. Kn\"{o}ll, W. Vogel, and D.-G. Welsch, J. Opt. Soc. Am. B {\bf 3},  1315
  (1986).

\bibitem{vogel}
W. Vogel and D.-G. Welsch, {\em Quantum Optics}, 3rd ed. (WILEY-VCH Verlag,
  ADDRESS, 2006).

\bibitem{breuer_book}
H.-P. Breuer and F. Petruccione, {\em The Theory of Open Quantum Systems}
  (Oxford University Press, ADDRESS, 2007).

\bibitem{PhysRevA.59.2306}
M.~W. Jack, M.~J. Collett, and D.~F. Walls, Phys. Rev. A {\bf 59},  2306
  (1999).

\bibitem{PhysRevLett.102.018303}
G. Bel and F.~L.~H. Brown, Phys. Rev. Lett. {\bf 102},  018303  (2009).

\bibitem{mukamel_book}
S. Mukamel, {\em Principles of Nonlinear Optical Spectroscopy} (Oxford
  University Press, ADDRESS, 1995).

\bibitem{shatokhin12b}
V. Shatokhin and T. Wellens, Phys. Rev. A {\bf 86},  043808  (2012).

\bibitem{PhysRevLett.96.100403}
M. Kiffner, J. Evers, and C.~H. Keitel, Phys. Rev. Lett. {\bf 96},  100403
  (2006).

\bibitem{PhysRevA.73.063814}
M. Kiffner, J. Evers, and C.~H. Keitel, Phys. Rev. A {\bf 73},  063814  (2006).

\bibitem{Cohen-Tannoudji79}
C. Cohen-Tannoudji and S. Reynaud, Phil. Trans. R. Soc. London A {\bf 293},
  223  (1979).

\bibitem{Armstrong:66}
J.~A. Armstrong, J. Opt. Soc. Am. {\bf 56},  1024  (1966).

\bibitem{1464-4266-2-2-317}
K. Joosten and G. Nienhuis, J. Opt. B: Quant. Semicl. Opt. {\bf 2},  158
  (2000).

\bibitem{Shatokhin01}
V.~N. Shatokhin and S.~Y. Kilin, Phys. Rev. A {\bf 63},  023803  (2001).

\bibitem{PhysRevLett.115.196402}
J.~C. L\'opez Carre\~no {\it et~al.}, Phys. Rev. Lett. {\bf 115},  196402
  (2015).

\end{thebibliography}

\end{document}